\documentclass[iop]{emulateapj}
\usepackage{graphicx}
\usepackage{amsmath}
\usepackage{afterpage}

\slugcomment{(Draft \today)}

\def\etal{\hbox{et al.}}

\gdef\ltsima{$\scriptscriptstyle \; \buildrel < \over \sim \;$}
\gdef\simlt{\lower.3ex\hbox{\ltsima}}

\gdef\gtsima{$\scriptscriptstyle \; \buildrel > \over \sim \;$}
\gdef\simgt{\lower.3ex\hbox{\gtsima}}

\gdef\about{\raise.3ex\hbox{$\scriptscriptstyle \sim $}}

\shortauthors{Kelson}

\shorttitle{How I Learned to Stop Worrying and Love the Central Limit Theorem}

\begin{document}

\title{Decoding the Star-Forming Main Sequence or:\break
How I Learned to Stop Worrying and Love the Central Limit Theorem}

\author{Daniel D. Kelson}
\affil{The Observatories of the
Carnegie Institution of Washington;
813 Santa Barbara St.;
Pasadena, CA 91101}

\begin{abstract}
Star-formation rates (SFR) of disk galaxies strongly correlate with stellar mass, with a small dispersion in specific
star-formation rate (SSFR) at fixed mass, $\sigma_{\log\text{SSFR}}\sim{0.3-0.4}$ dex. With such small scatter this
``main sequence of star-formation'' (SFMS) has been interpreted as deterministic and fundamental to galaxy evolution.
Here we demonstrate that the SFMS is a simple consequence of the central limit theorem.
Our derivation begins by approximating the {\it in situ\/} stellar mass growth of galaxies as a stochastic process,
much like a random walk (where the expectation of SFR at any timestep is equal to the SFR at the previous timestep).
We then derive expectation values for the median SSFR of
star-forming disks and their scatter over time. We generalize the results to encompass stochastic changes in SFR
that are not fully independent of each other but are correlated over time. For fair, unbiased
samples of (disk) galaxies, we derive an expectation that $\langle\text{SSFR}\rangle$ should be independent of mass,
decline as $1/T$, and have a relative scatter that is also independent of mass and time. The derived SFMS and its
evolution matches published data over $0\le z \le 10$ with sufficient accuracy to constrain cosmological parameters.
This framework reproduces, with no prior inputs, several important observables, including: the scatter in SSFR at fixed
mass; the star-formation histories of nearby dwarf galaxies and the Milky Way; and the scatter in the Tully-Fisher
relation. The evolution of the stellar mass function is less well reproduced but we discuss avenues for generalizing the
framework to include other sources of stellar mass such as mergers and accretion. The predicted dispersion in
SSFR has consequences for the classification of quiescent galaxies, as such galaxies have heterogeneous formation
histories, and many may only be temporarily diminished in their star-formation activity. The implied dispersion in SFHs,
and the SFMS's insensitivity to timescales of stochasticity, thus substantially limits the ability to connect massive
galaxies to their average progenitors over long cosmic baselines.
\end{abstract}


\keywords{
galaxies: evolution ---
galaxies: stellar content
}

\section{Introduction}
\label{sec:intro}

Our understanding of the present-day distributions of galaxies and their properties has largely
focused on the accounting of stellar mass back through time.
Such work is routine, with snapshots of the distribution of mass in both active and passive galaxies from $z\sim
0$ \citep[e.g.][]{cole2001,bell2003,moustakas2013} to $z=2$ \citep{tomczak2014}, and galaxy mass
functions back to times when the Universe was under a Gyr old \citep{gonzalez2011}.
Out of these efforts we see a striking tenfold increase in the numbers of quiescent $M^*$ galaxies since $z=2$
\citep{muzzin2013}, with a twofold increase since just $z=1$ \citep[e.g.][]{brown2007}.

Despite such evolution in the numbers of normal and massive galaxies, the processes that
produced such growth remain elusive. The conundrum is compounded by the relative invariance in the
shape the passive galaxy mass function \citep[e.g.][]{muzzin2013,kelson2014a}.
How are so many passive galaxies created over such a broad range of stellar mass at relatively
equal rates during the past 10 Gyr while preserving the overall distribution of ellipticities
\citep{holden2012}? And why do the quiescent and star-forming mass functions have the
same slope at low masses, with a quiescent galaxy fraction of 15\%-20\% that stays relatively constant with time
\citep{muzzin2013}?

Furthermore, the processes that make such tremendous numbers of massive galaxies quiescent over the past 10 Gyr have
also served to increase their apparent sizes \citep{pvd2008,pvd2010}, with some attendant morphological
transformations \citep{vdw2011}. Accounting for this size evolution has required invoking minor {\it and\/} dry
mergers \citep{bezanson2009}, as there are apparently too few substantial, or potentially substantial mergers
\citep{williams2011} to get the job done despite the well-known correlations between young stellar populations and fine
structures and tidal features in early-type galaxies \citep[e.g.][]{schweizer1990,schweizer1992}.
The evolving morphological mix of quiescent galaxies since $z\sim 2$
surely indicates that both disks and bulges play crucial roles \citep{bell2012}, that the processes that both grow
galaxies and lead to increased quiescence are not decoupled.

Snapshots of stellar mass alone, however, have not uncovered the mechanisms that gave rise to such an abundance of
quiescent galaxies at late times. The distribution of ongoing star-formation rates (SFR) at fixed stellar mass has
been taken as a crucial piece of the puzzle; such data offer the only direct window into the key process by which
galaxies grow. Locally there exists a strong correlation between SFR and stellar mass, with
a slope near unity and
$0.3-0.4$ dex of scatter \cite[e.g.][]{brinchmann2004,salim2007}. At modest masses, $M>5\times{10^{9}}M_\odot$, the
slope changes, such that SSFR declines mildly with mass, and such a shape is observed back to $z\sim 2$
\citep{karim2011,whitaker2012,sobral2014}.

The relative constancy of SSFR is commonly used to infer that star-formation proceeds with similar efficiency in all
galaxies. After all, baryonic mass must scale roughly with halo mass, and since gas mass roughly scales with baryonic mass
the constant SSFR must
imply a universal efficiency with which galaxies convert gas into stars. At high masses the change in slope implies
diminished efficiency, though recent work by \cite{abramson2014} shows that at least some of
the anticorrelation between SSFR and galaxy mass arises because massive galaxies tend to have higher bulge mass
fractions, on average, and bulges tend not to form stars in great numbers.

The fact that bulges do not participate meaningfully in the formation of stars compared to disks means
that one should decouple the processes of star-formation and bulge growth from the SFMS. If one wants to probe
the extent to which
star-formation efficiency is a property of galaxy disks, then normalizing SFR by the total stellar mass
of galaxies is unhelpful. Using a large sample of galaxies in the SDSS,
\cite{abramson2014} deduced the stellar masses of galaxy disks and constructed a new SFMS using SFR/$M_\text{disk}$.
They found that the presence of bulge mass had accounted for $-0.3$ dex/dex of the slope of the SFMS at high masses.
If bulge formation is a process that is distinct from the {\it in situ\/} stellar mass growth of galaxies, then
the steeper slopes derived earlier for the high-mass end of the SFMS was providing a skewed picture of the processes
that quench galaxies. Naturally one wants to understand both the histories of star-formation and
mass assembly, as the latter process serves to increase the denominator of SSFR with potentially negligible
impact on the numerator. \cite{abramson2014} found that there was significantly less dependence of the star-formation
efficiency of galaxy disks on galaxy mass than one would have deduced by assuming galaxies were monolithic structures.

Despite such trivialities as galactic structure, the redshift dependence of the SFMS had been coupled with the
evolving stellar mass function to provide functional frameworks to describe the growth and quenching of galaxies over
time \citep{peng2010,peng2012,behroozi2013}. These physical interpretations for scaling relations such as the SFMS are
on one end of a spectrum of even more informative, yet deterministic, schemes for describing the growth of galaxies in
cosmological contexts \citep{bower2006,benson2012,benson2014}.

Here we consider the SFMS from a different standpoint, recognizing that the growth of stellar mass occurs within
a tempestuous context, over long and short timescales, and a myriad of environments. As such, we attempt to
treat the {\it in situ\/} growth of stellar mass, i.e. star-formation, as a stochastic process. Stochastic processes
have strict definitions in the mathematical and statistical literature that are different than what astronomers
normally assume when loosely using the word ``stochastic.'' A stochastic process is
not a random process, but it is defined by a formal expectation that on average they do
{\it not\/} change from one moment to the next. Under this formalism, the central limit theorem produces a
correlation between SFR and stellar mass strikingly similar to the one observed.

Finding that the SFMS is a statistical artifact may run counter to one's intuition --- 
how can a purely mathematical derivation of a distribution of SSFRs properly reflect the range of astrophysical
phenomena that grow galaxies, as well as the potentially broad range of timescales that govern stochasticity in
a cosmological context? One very specific answer calls to our intuition as astronomers: if a galaxy of a given mass
and star-formation rate today probably didn't have a very different star-formation rate a short time ago, and then again
a short time before that, and before that, then there's a limited set of histories that could have resulted in that
galaxy having that mass at the present epoch. Any significant perturbation to its star-formation history (SFH) would
have produced a galaxy with a different mass. It is this normalization of star-formation by stellar mass that is the key
to seeing the SFMS as a scale-free consequence of the stochastic processes that grew galaxies over time.

From this point of view, we derive
a statistical description of the ensemble of star-forming galaxies using a minimum of deterministic rules.
For much of the text we treat the ensemble of galaxies as an ensemble of systems that form stars. Because the bulk of
star-formation occurs not in bulges, we assume our formalism pertains specifically to galaxy disks, or at least the
general class of star-forming disk galaxies. However, this distinction is made more out of convenience so that the
methodology can be generalized to encompass other sources of stellar mass, such as merging and accretion. Such
processes also bring fuel for star-formation, but the resulting growth of stellar mass out of that gas is
observed through measurements of {\it in situ\/} SFRs and as such is largely accounted for in our derivation of the SFMS.
Such details are deferred until later, as we have not yet left the Introduction.

In \S \ref{sec:process} stochastic processes are defined for the purposes of modeling stellar mass growth, and the
central limit theorem is applied in order to derive the long-term behavior in distribution. In \S \ref{sec:fBm} we generalize
the results to histories with long-term correlations between stochastic events. Within this formalism we discuss a number
of immediate consequences for the general distributions of star-forming galaxies in \S \ref{sec:predict}. In \S
\ref{sec:mf} we extend the mathematical consequences of stochasticity to a general discussion of stellar mass growth
over cosmic time. \S \ref{sec:ambig} we provide a few examples where stochasticity limits our ability to interpret
observations. Finally in \S \ref{sec:conclusions} we summarize the key points. All equations derived below have been
verified with numerical experiments.


\section{Expectations for Stochastic Star-Formation}
\label{sec:process}

\subsection{Operational Definition}

Stochastic processes are not anarchic, but remain constant {\it on average\/} from one timestep to the next, much
like a random walk or gambling on a coin toss.
An analogy: go measure star-formation rates for a sample of galaxies. What will their SFRs be
in the near future? Though the SFRs of individual galaxies may have experienced small changes, on average things will
not look very different.

So let us consider a process, $S$, as the amount of stellar mass formed over time interval $t$.
\begin{eqnarray}
S_t&=\label{eq:dmdt}&{M_{t+1}-M_t\over{\Delta{\mathcal{T}}_t}}\Delta{\mathcal{T}}_t
\end{eqnarray}
where $M_t$ is the mass accumulated up to the interval $t$ in time $\mathcal{T}$ by {\it in situ} mass growth. We
will assume discrete timesteps for the moment, with no loss of generality.

If $S_t$ is a stochastic process, then at time $t+1$ there is an expectation that $S_{t+1}$ is,
on average, the same as $S_t$:
\begin{eqnarray}
\mathcal{E}[S_{t+1}]=S_t\label{eq:stationary}
\end{eqnarray}
In other words, 10 years from now we expect the Milky Way to be forming stars at the same rate that it does today,
and the same should be true 100 years from now, and 10 million years from now, or even 100 million years from now.
But over time there are probabilities that the star-formation rate may go up a little or go down a little. So
from one timestep to the next there are changes to $S$, which we define as $X$:
\begin{eqnarray}
X_{t+1} = S_{t+1}-S_{t}
\end{eqnarray}
At this point in the derivation the astronomer should note that physics is buried here, in $X$, where
a broad range of astrophysical processes can serve to increase or decrease $S$, the rate at which a galaxy grows.
These ``stochastic differences'' in $S$ are independent random numbers drawn from distributions
centered on zero, because that is the definition of a stochastic process, and because there are astrophysical
reasons why the rates of growth for galaxies may increase or decrease. We will revisit this assumption later as
we generalize the derivations for correlations among the stochastic differences, and for a more general picture
of galaxy assembly. For now we assume these stochastic differences are all independent, so that
at each timestep, the random variables $X$ are also be drawn
from distributions with potentially time-dependent variances:
\begin{eqnarray}
\text{Var}[S_{t+1}-S_t]=\sigma^2_{t+1}\label{eq:variance}
\end{eqnarray}

We can rewrite the stellar mass formed at time $t$ as the sum of the stochastic differences:
\begin{eqnarray}
S_t&=\label{eq:stoch}&(S_t-S_{t-1}) + (S_{t-1}-S_{t-2}) + \ldots +S_0\\
&=&\sum_{i=1}^{t} X_i
\end{eqnarray}
where we set $S_0=0$ for simplicity.

By definition the accumulated stellar mass growth at time $T+1$ is
\begin{eqnarray}
M_{T+1}&=&\sum_{t=1}^TS_t\\
&=&\sum_{t=1}^T\sum_{i=1}^{t}X_i
\end{eqnarray}

Note that the SFMS is an observed correlation of $S_T$ vs $M_T$, and
we have now shown that it is really
between $\sum_{t=1}^{T} X_t$ and $\sum_{t=1}^{T-1} \sum_{i=1}^t X_t$. This recasting is important as the variances in
$X$ will drive both $S$ and $M$ from only the very few assumptions made so far. We now
proceed to derive long-term expectation values for both $S$ and $M$, as well as an expected distribution of $S/M$
($\equiv\text{SSFR}$; specific star-formation rate).

\subsection{The Martingale Central Limit Theorem}
\label{sec:variance}

If the stochastic differences, $X_t$, are independent, random variables centered on zero --- with equal probability of
being positive or negative --- $S$ is called a martingale, and $X$ are called martingale differences. Martingales were
originally defined in the context of gambling as ``fair'' games, where, at any time or step,
$S$ (your winnings) may go up or down with equal chance, by an amount $X$,
{\it independent of earlier events\/}. There exist central limit theorems that describe their long-term behavior in
distribution \citep{hall1980}, briefly summarized here.

We introduced stellar mass growth as a stochastic process, and continue by defining the
variance in $S_T$:
\begin{eqnarray}
\text{Var}[S_T]&=\label{eq:var2}&\mathcal{E}[S_T^2]-(\mathcal{E}[S_T])^2
\end{eqnarray}
When the stochastic differences are independent, the variance in $S_T$, is simply the sum of the variances in the
stochastic differences leading up to that time. Because $S$ is stationary,
\begin{eqnarray}
\text{Var}[S_T]\label{eq:clt1}&=&\sum^{T}_{t=1}X_t^2=\sum^{T}_{t=1}\sigma_t^2
\end{eqnarray}
This result appears similar to Brownian motion, itself a simple stochastic process, though here we have allowed, for
example, the variances to also be random, and for the distributions of stochastic differences to be non-Gaussian.
By doing so, we attempt a derivation with as few assumptions about stellar mass growth as possible.

Given $N$ galaxies with their stochastic histories, we define a normalization for object $n$
that is equivalent to an expected rms fluctuation in $S_{n,T}$:
\begin{eqnarray}
\sigma_{n,T}&=\label{eq:norm}&\biggl[\biggl({1\over{T}}\biggr)\sum_{t=1}^T\sigma_{n,t}^2\biggr]^{1/2}
\end{eqnarray}

The central limit theorem states that the distribution of $S_T$ will be a Gaussian centered on zero with a standard
deviation of unity.
In other words, $S_{n,T}$, where $n\in\{1,2,\ldots,N\}$,
converges {\it in distribution\/} to the
normal distribution, $\mathcal{N}(0,1)$ \citep{hall1980}, when $S$ is properly normalized:
\begin{eqnarray}
{S_{n,T}\over{T^{1/2}\sigma_{n,T}}}&=&{1\over{T^{1/2}\sigma_{n,T}}}\sum_{t=1}^{T}X_{n,i}\\
{S_T\over{T^{1/2}\sigma_T}}&\xrightarrow{d}&\mathcal{N}(0,1)\label{eq:gauss1}
\end{eqnarray}

What we see already in Equation \ref{eq:gauss1} is that given some astrophysical variance $\sigma^2_{n,T}$, known or
measured {\it a priori\/} for a distribution of galaxies or proto-galaxies, we can already calculate an expected
variance in stellar mass growth at any epoch $T$. But before we apply such a probability distribution to star-formation,
we now impose nonnegativity on $S_t$.

\subsection{Nonnegativity}

Galaxies grow stars at nonnegative rates so our solutions for the distributions of $S$ must be bounded by $S_t\ge{0}$.
One approach is to approximate {\it in situ} stellar mass growth in a manner similar to geometric Brownian motion,
where one models distributions of $\ln S$, and the stochastic differences $X$ are changes in the logarithm of SFR.
Doing so would undoubtedly ensure positivity, but such an approach would preclude the possibility that (some) galaxies
may stochastically be driven to star formation rates of zero.

By adopting a boundary condition of $S_t\ge{0}$, we obtain a probability density for $S_T$ that is not a symmetric
Gaussian centered on zero, as shown in Equation \ref{eq:gauss1}, but is simply the nonnegative side of the Gaussian:
\begin{eqnarray}
P\bigl[{S_T\over T^{1/2}\sigma}<x\bigr]\label{eq:pnn}&=&\biggl({2\over\pi}\biggr)^{1/2}\int_{0}^{x}e^{-{1\over 2}z^2}\text{d}z
\end{eqnarray}
We obtain this result because technically $S$ is no long a martingale but a submartingale. Loosely a submartingale is
defined by
\begin{eqnarray}
\mathcal{E}[S_{t+1}]\ge S_t\label{eq:sub}
\end{eqnarray}
For example, when a galaxy's star-formation rate has been stochastically driven to $S_t=0$, the SFR at the next interval
cannot be driven to lower rate and can only go up. Thus $S$ is no longer a ``fair game.''
With submartingales there are several tricks for deriving their limiting behavior, but in this case it simplest to use
the theorem that every submartingale can be written as the absolute value of a martingale \citep{gilat1977}. In doing so
one derives the above result (with numerical experiments verifying this behavior).

The expectation values for $S_t$ and $M_t$ over $1\le{t}\le{T}$ are
\begin{eqnarray}
\mathcal{E}\bigl[{S_t\over T^{1/2}\sigma}\bigr]&\label{eq:st}=&\biggl({2\over{\pi}}\biggr)^{1/2}\biggl({t\over{T}}\biggr)^{1/2}\\
\mathcal{E}\bigl[{M_t\over T^{1/2}\sigma}\bigr]&\label{eq:mt}=&{2\over{3}}T\biggl({2\over{\pi}}\biggr)^{1/2}\biggl({t\over{T}}\biggr)^{3/2}
\end{eqnarray}
with the variance:
\begin{eqnarray}
\text{Var}\bigl[{S_t\over T^{1/2}\sigma}\bigr]\label{eq:clt2}&=&{1\over{2}}\mathcal{E}\bigl[{S_t\over T^{1/2}\sigma}\bigr]^2
\end{eqnarray}
where $T^{1/2}\sigma$ is just the normalization for the central limit theorem (Equation \ref{eq:gauss1}).

For $S_T/M_T(\equiv\text{SSFR})$ the expectations are, including the expectation for the standard deviation:
\begin{eqnarray}
\mathcal{E}\bigl[{S_T\over M_T}\bigr]\label{eq:clt4}&=&{3\over{2T}}\\
\text{Sig}\bigl[{S_T\over M_T}\bigr]\label{eq:clt5}&=&{1\over{\sqrt{2}}}\mathcal{E}\bigl[{S_T\over M_T}\bigr]\\
\text{Sig}\bigl[\ln {S_T\over M_T}\bigr]&\approx&{1\over{\sqrt{2}}}\\
\text{Sig}\bigl[\log {S_T\over M_T}\bigr]&\approx&0.3\hbox{ dex}
\end{eqnarray}
What these derivations show is that under the assumptions outlined above, we would expect SSFR to (1) be
mass-independent, (2) fall like $1/T$, and (3) have a {\it relative\/} dispersion that should be independent of mass and
time, at a level of $\sim 0.3$ dex. We have neither made, nor required, any assumptions about the
star-formation efficiencies of galaxies to reach these results.

These expectation values can be directly contrasted to the case where there are zero stochastic changes to
star-formation, as such a case would produce $\mathcal{E}[S_T/M_T]=1/T$, and
$\text{Sig}[S_T/M_T]=0$. Such was the case explored by the models of \cite{peng2010,peng2012}. Random
noise in the star-formation histories leads to an increase in the median SSFR by 50\%, but leaves the correlation between
SFR and mass in tact with fairly small scatter.

Before we can directly compare these expectations to the observed SFMS, as well as other data, we will
generalize the derivation in \S \ref{sec:fBm} for expected short- and long-term correlations between the random,
stochastic changes in star-formation. And before doing that, we first have two digressions: one on timescales, and another on
how notions of star-formation efficiency might fit into this framework.

\subsection{A Brief Digression on Timescales}
\label{sec:time}

Until now we have kept the derivations simple by adopting constant integer intervals for timesteps. The reader may
wonder how our derivations may depend on the timescales over which galaxies experience stochastic changes to their
stellar mass growth. Equation \ref{eq:dmdt} explicitly allows for arbitrary intervals in time, such that
\begin{eqnarray}
S_t&=\label{eq:dmdt2}&\langle{\dot{M}}\rangle_t\Delta{\mathcal{T}}_t
\end{eqnarray}
where $\langle{\dot{M}}\rangle_t$ is the mean star-formation rate during the $t$th interval
in time. Thus
\begin{eqnarray}
\langle{\dot{M}}\rangle_T\Delta{\mathcal{T}}_T&=\label{eq:dmdt3}&\sum_{t=1}^{T}
\bigl(\langle{\dot{M}}\rangle_t\Delta{\mathcal{T}}_t - \langle{\dot{M}}\rangle_{t-1}\Delta{\mathcal{T}}_{t-1}\bigr)
\end{eqnarray}
and
\begin{eqnarray}
M_{T+1}&=&\sum_{t=1}^{T} \langle{\dot{M}}\rangle_t\Delta{\mathcal{T}}_t\\
&=\label{eq:dmdt4}& \sum_{t=1}^{T} \sum_{i=1}^{t}
\bigl(\langle{\dot{M}}\rangle_i\Delta{\mathcal{T}}_i - \langle{\dot{M}}\rangle_{i-1}\Delta{\mathcal{T}}_{i-1}\bigr)
\end{eqnarray}

While it is not strictly necessary to adopt constant intervals in time (for a given galaxy), we do so below to simplify
the presentation. The equations for randomly varying timesteps is a little more involved than is needed here, though we
explore the consequences of random timescales later in \S \ref{sec:time2}.

So let us now write
\begin{eqnarray}
\langle{\dot{M}}\rangle_T&=&\sum_{t=1}^{T}
\bigl(\langle{\dot{M}}\rangle_t
- \langle{\dot{M}}\rangle_{t-1} \bigr)
\end{eqnarray}
and
\begin{eqnarray}
M_{T+1}&=&\Delta{\mathcal{T}} \sum_{t=1}^{T}
\langle{\dot{M}}\rangle_t\\
&=&\Delta{\mathcal{T}} \sum_{t=1}^{T} \sum_{i=1}^{t}
\bigl(\langle{\dot{M}}\rangle_i 
- \langle{\dot{M}}\rangle_{i-1}  \bigr)
\end{eqnarray}

The martingale differences
\begin{eqnarray}
X'_t = \langle{\dot{M}}\rangle_t- \langle{\dot{M}}\rangle_{t-1}
\end{eqnarray}
are drawn from distributions with ${\sigma'_t}^2$. After normalizing as in Equation \ref{eq:norm} and rearranging,
we obtain
\begin{eqnarray}
\mathcal{E}[{\langle\dot{M}\rangle}_T]&=\label{eq:stp}&\sigma'\biggl({2\over{\pi}}\biggr)^{1/2}{T}^{1/2}\\
\mathcal{E}[M_T]&=\label{eq:mtp}&\Delta{\mathcal{T}}\sigma'{2\over{3}}\biggl({2\over{\pi}}\biggr)^{1/2}{T}^{3/2}
\end{eqnarray}

Using these results, we therefore derive an expectation value for SSFR for the ensemble of disk galaxies whose
stochastic changes in SFR occur on timescales of $\Delta{\mathcal{T}}$:
\begin{eqnarray}
\mathcal{E}[{\langle\dot{M}\rangle_T}/M_T]&=\label{eq:time2}&{3\over{2T\Delta{\mathcal{T}}}}
\end{eqnarray}

And since at any given epoch $T\Delta{\mathcal{T}}$ is just the time since the beginning of star-formation, we recover
the earlier result that $\mathcal{E}[S/M]=3/(2\mathcal{T})$. Put more concretely: galaxy disks that experience $10$ major
stochastic changes to their stellar mass growth over a random distribution of timescales since $z=20$ will have the same
distribution of SSFRs as galaxies that experience $100$ major episodes over the same span of cosmic time.

In summary, Equation \ref{eq:time2} adds another key point to the conclusions drawn earlier for the distribution of
SSFRs for fair, representative samples of star-forming disk galaxies. Not only is
$\mathcal{E}[{\langle\dot{M}\rangle_t}/M_t]$ independent of mass, and not only does it fall by $\mathcal{T}^{-1}$, and
not only is its relative scatter independent of mass and time, but these expectations are independent of the timescales
over which stochastic changes in star-formation occur. In other words, {\it almost\/} no information is encoded within
diagrams of SSFR vs stellar mass regarding the timescales of the processes that bounce galaxies up or down or left and
right. Star-formation rate indicators that probe a range of timescales within a given galaxy will, of course, provide
useful information on formation histories, but the derivation of individual histories from the SFMS is
mathematically ill-defined.

At least as derived so far, and we are not yet finished, the SFMS appears to be as fundamental a scaling relation as one
might ever find in nature, arising not from any astrophysically deterministic laws of galaxy
evolution, but seemingly arising from one of the most profound mathematical ones, the central limit theorem.

\subsection{A Brief Note on Star-Formation Efficiency}
\label{sec:eff}

We have one more brief aside before generalizing our derivations for correlated  stochastic changes in SFR, followed by
a short discussion of observations related to these statistical expectations. In that context we will also
discuss general extensions to our formalism to include {\it ex situ\/} stellar mass growth, and variable timescales
of stochasticity.

Equations \ref{eq:stationary} through Equations \ref{eq:time2} were written assuming $S$ as increments of {\it in
situ\/} stellar mass growth. However, one may write
\begin{eqnarray}
S_t=\epsilon_t{F_t}
\end{eqnarray}
where $F_t$ is the fuel available for making new stellar mass at time $t$ and $\epsilon$ is the efficiency with which
fuel is converted to new stars.
Such notation may explicitly incorporate stochastic changes in efficiency over time, as well as changes in the
availability of cold gas to form new stars, but the derivation of the SFMS continues unaltered.

While we are not finished deriving what will be the final form of the SFMS for disk galaxies, we have
already concluded that the observed correlation between SFR and stellar mass for disk galaxies is independent
of mass, time, or even timescales, and it is equally important to stress that the existence of the SFMS 
does not imply that galaxies form stars with uniform, or nearly uniform efficiency.

It may prove interesting in the future to employ $\epsilon$ and $F$, if only to use observations outside of the
SFMS to constrain the astrophysical processes and events that drive stochastic changes in SFR. We will comment more later, 
but no matter the cause, so long as the stochastic differences are random and stationary, no amount of tinkering with
astrophysical underpinnings will move the expectation values for star-forming disks.
All astrophysics is subsumed within $\sigma$, defined by Equation \ref{eq:norm}. Recasting $S$ as a combination of
star-formation efficiency and fuel supply only changes the philosophical underpinnings of $\sigma$ but does not
prevent it from cancelling out when dividing $S$ by $M$.

\begin{figure*}[t]
\centerline{\includegraphics[width=6.5in]{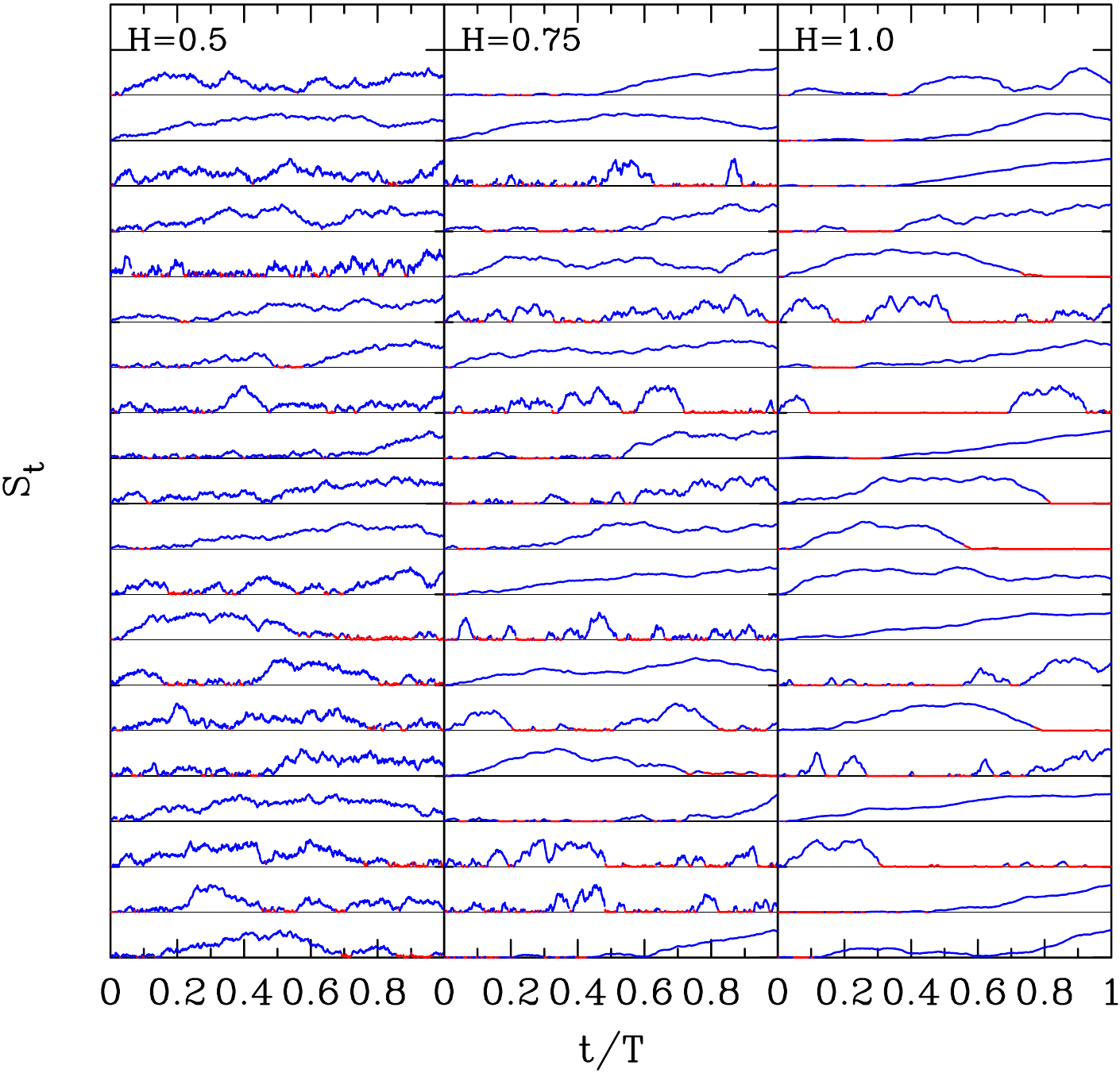}}
\caption{Randomly generated paths of stochastic histories with variable long-term covariance. The case where stochastic
changes in $S$ are independent is shown on the left, with $H=0.5$. Progressively towards the right are models with
increasing levels of covariance, with a maximum of $H=1$. Segments of these histories are color coded red when
$S_t\le{M_t/(4t)}$, and blue when $S_t$ is larger.
\label{fig:sfhplot}}
\end{figure*}

\begin{figure*}[t]
\centerline{\includegraphics[width=7.0in]{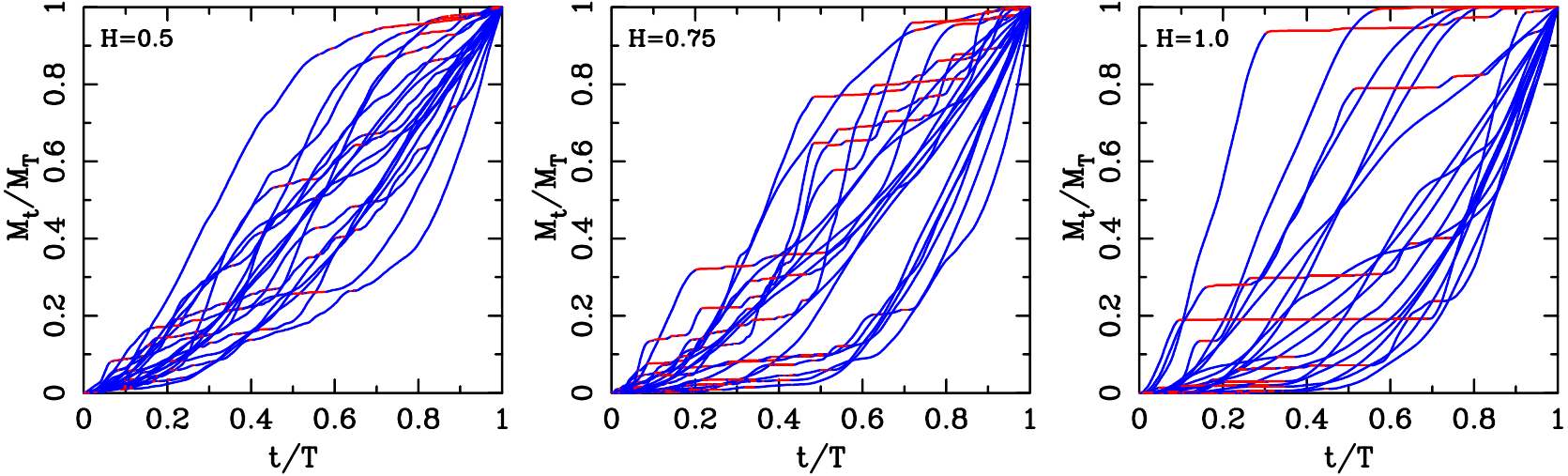}}
\caption{The growth histories for the paths shown in Figure \ref{fig:sfhplot}.
Though there is a significantly greater diversity of star-formation histories as $H$ increases,
the growth histories from smaller values of $H$ appear to be subsumed into a larger, more general set of
curves. The level of correlation among stochastic changes to star-formation over a Hubble time defines the
possible range of histories galaxies may have experienced to reach a particular stellar mass at a given
epoch.
\label{fig:sfhplot2}}
\end{figure*}

\section{Expectations for Covariant Stochasticity}
\label{sec:fBm}

The long (infinite) reach of gravity, plus the primordial power spectrum, correlates events over a broad range of
time-scales. Unfortunately the derivation in \S \ref{sec:process} assumed that stochastic changes in star-formation
were independent and uncorrelated. When events are correlated Equation \ref{eq:stoch} must be modified
\begin{eqnarray}
S_T&=\label{eq:covstoch}&\sum_{t=1}^{T}\sum_{i=0}^{m}c_{t,t-i}X_{t-i}
\end{eqnarray}
where $m$ denotes a maximum number of covariant timesteps, and $c_{t,t-i}$ are the unknown correlation
coefficients between the stochastic changes in $S$ at times $t$ and $t-i$.

Because $m$ describes the correlation timescale over which changes in $S$ are arbitrarily covariant, these changes are
referred to as ``$m$-dependent random variables.'' Fortunately such random variables also obey central limit theorems
\cite[e.g][]{romano2000}. Furthermore, for the form of correlations expected in a cosmological context, the sums of
$m$-dependent random variables converge {\it in distribution\/} to fractional Brownian motions
\citep{pipiras2000}. In \S \ref{sec:process} our derivations converged to nonnegative Brownian motions.

Fractional Brownian motions \citep[e.g.][]{mandelbrot1968,mandelbrot1969} are ``fractional'' in the sense that the
stochastic changes in $S$ are not fully independent of each other. Thus, fBm is a generalization
of the formalism derived earlier --- in which we derived expectations for purely random changes in SFR --- such that the
random changes are smoothed on all timescales. This smoothing is controlled by $H$, the Hurst parameter, where
$0\le{H}\le{1}$ \citep{mandelbrot1968}. When $H=0.5$ there is no covariance between the stochastic changes in $S$ and
the derivations in \S \ref{sec:process} are recovered exactly.

In contrast, histories with $H<0.5$ show rapid variability, where increases in $S$ tend to be followed rapidly by
stochastic decreases, ever trying to revert $S$ to the mean. In the limit of $H=0$, the solutions are equivalent to
constant star-formation, with $\text{SSFR}=1/T$ and zero intrinsic scatter. When $H>0.5$, there are long-term trends,
since increases tend to be followed by more increases, and decreases tend to be followed by more decreases. The larger
$H$ is, the more the implied histories are marked by extremes of feasts and famines, as one might expect in a universe
of hierarchical galaxy and structure formation.

It is remarkable and convenient that the unknown, random distributions of stochastic
changes in star-formation {\it and the unknown, random covariances between them over arbitrarily long timescales\/}
may be reduced, in distribution, to a single number: $H$. Normalizing by $T^{1/2}\sigma$ as above, fBm has
well-behaved expectations for $S_t$, with simple dependencies on $H$  \citep{mandelbrot1968}. In the coming sections we
will compare these long-term expectation values with a range of data but first we show example fBm paths for $S$.

\subsection{Examples}

Figure \ref{fig:sfhplot} shows example (normalized) nonnegative fBm realizations with $T=1000$ timesteps and
three levels of covariance $H\in\{0.5,0.75,1.0\}$.
Even for the case of uncorrelated changes in SFR, with $H=0.5$, the median absolute relative change in $S$ from one
timestep to the next is $\sim{3}\%$. The level of fractional change scales as $T^{-1/2}$ so when galaxies experience
only $T=10$ major episodes over a Hubble time, the typical fractional change will be a factor of 10 higher. For those
galaxies that experience $10^5$ stochastic changes, then the typical fractional change will be $\ll 1\%$.
Over any number of timesteps within the same amount of cosmic time, $\mathcal{T}$,
$S$ slowly wanders and the ratio of $S/M$ follows the well-defined distribution derived earlier.

When the stochastic differences $X$ are correlated over time, the relative changes in $S$ from timestep to timestep are
much smaller. With $T=1000$ timesteps and $H=1$ the median absolute relative change in $S$ is $\sim{0.25\%}$. In the
figure the paths are color coded red when $S_t$ is less than a quarter the lifetime average, and blue when $S_t$ is
greater. Such demarcation crudely separates when each aggregate ``stellar population'' might appear red or blue. The
normalized growth histories for these paths are shown in Figure \ref{fig:sfhplot2}, color coded similarly.

It is important to note that the formal rules of stochastic processes have led to suites of growth histories, depending
on the level at which stochastic changes to SFR are correlated over time. The central limit theorem provides probability
densities for these histories such that an aggregate distribution of SFHs can be constructed {\it a priori\/}, based on
the amount of time a galaxy had to grow to its given mass. However, nothing in the rules, so far, has limited which
histories may be unphysical or potentially disallowed. In theory these are the distributions for unbiased ensembles,
where SFRs have an equal chance of increasing or decreasing at every (unknown) timestep. Astrophysical simulations may
serve to constrain these sets to better represent the cosmological distribution of histories that real star-forming
disks experience. We introduce generalizations to this formalism later in the text to help account for astrophysically
interesting constraints, but proceed under the assumption, for now, that these general distributions of SFHs represent
the range of possible histories for the Universe's ensemble of galaxies as they evolve through a broad range of
dynamically changing galaxy environments over cosmic time.

\subsection{Expectation Values}

As $H$ increases fBm goes from producing a narrow range of growth curves to progressively greater and
greater diversity in growth histories and $S$ becomes increasingly smooth. Even with such a diversity of
``star-formation histories,'' limit theory can describe their distributions.

In distribution, the ratio of $S_t/M_t$ for arbitrary $H$ can be derived from the expectation values of $S_t$ and $M_t$.
These are different than the classic fBm expectations in \cite{mandelbrot1968} because of the boundary
$S\ge 0$:
\begin{eqnarray}
\mathcal{E}\bigl[{S_t\over T^{1/2}\sigma}\bigr]&=&\biggl({2\over{\pi}}\biggr)^{1/2}
\biggl({T^{H-1/2}\over{2H}}\biggr)\biggl({t\over{T}}\biggr)^H\\
\mathcal{E}\bigl[{M_t\over T^{1/2}\sigma}\bigr]&=&\biggl({2\over{\pi}}\biggr)^{1/2}
\biggl[{T^{H+1/2}\over{2(1+H)H}}\biggr]\biggl({t\over{T}}\biggr)^{(1+H)}\\
\text{Var}\bigl[{S_t\over T^{1/2}\sigma}\bigr]&=&H\mathcal{E}\bigl[{S_t\over T^{1/2}\sigma}\bigr]^2
\end{eqnarray}
These scale-free forms can be simplified to
\begin{eqnarray}
\mathcal{E}\bigl[S_t\bigr]&\label{eq:sth}=&\sigma\biggl({2\over{\pi}}\biggr)^{1/2}\biggl({t^H\over{2H}}\biggr)\\
\mathcal{E}\bigl[M_t\bigr]&\label{eq:mth}=&\sigma\biggl({2\over{\pi}}\biggr)^{1/2}\biggl[{t^{H+1}\over{2(1+H)H}}\biggr]\\
\text{Var}\bigl[S_t\bigr]\label{eq:clt2h}&=&H\mathcal{E}\bigl[S_t\bigr]^2
\end{eqnarray}
where, once again, astrophysics has been subsumed in $\sigma$, though, now $H$ also represents long-term
astrophysical and cosmological effects through its control of the long-term correlations of stochastic events.
Given $H$, these expectation values define generalized versions of the star-forming main sequence and its scatter:
\begin{eqnarray}
\mathcal{E}[S_t/M_t]\label{eq:clt4h}&=&(H+1)\over{t}\\
\text{Sig}[S_t/M_t]\label{eq:clt5h}&=&{H^{1/2}}\mathcal{E}[S_t/M_t]
\end{eqnarray}
Inferring the distribution in $\log\text{SSFR}$ from these equations directly is not straightforward. Using numerical
experiments, we find the intrinsic distribution in $S/M$ is not quite Gaussian and not quite lognormal.
When $S/M>\mathcal{E}[S/M]$ the distribution of
$(S/M-\mathcal{E}[S/M])/(H^{1/2}\mathcal{E}[S/M])$ appears to be lognormal, and
when $S/M<\mathcal{E}[S/M]$ the distribution appears normal down to $S/M=0$.
And, of course, galaxies preselected as star-forming from any survey naturally
excludes those galaxies at $S/M=0$, and objects with low SSFR, depending on the nature of the selection.
More critically, these numerical experiments indicate that the expectation value $\mathcal{E}[S/M]$ is equivalent to the
median SSFR, not the mean (except when $H\equiv 0$).

Thus we now have derived the specific
value and evolution of ``the flat part'' of the SFMS, where SSFR is roughly constant with mass. 
These retrodictions should be valid for bulgeless galaxies --- so predominantly low-mass at late times and both low- and high-mass
galaxies at earlier times --- where all of the observed $M_T$ can be associated with {\it in situ\/} mass growth
\citep{abramson2014}. We discuss departures from this assumption in a later section within a broader discussion of how
distributions of galaxies may deviate from these simple expectations.

Several interesting consequences of Equations \ref{eq:clt4h} and \ref{eq:clt5h} are discussed in the next section.

\begin{figure*}[ht]
\centerline{\includegraphics[width=5.5in]{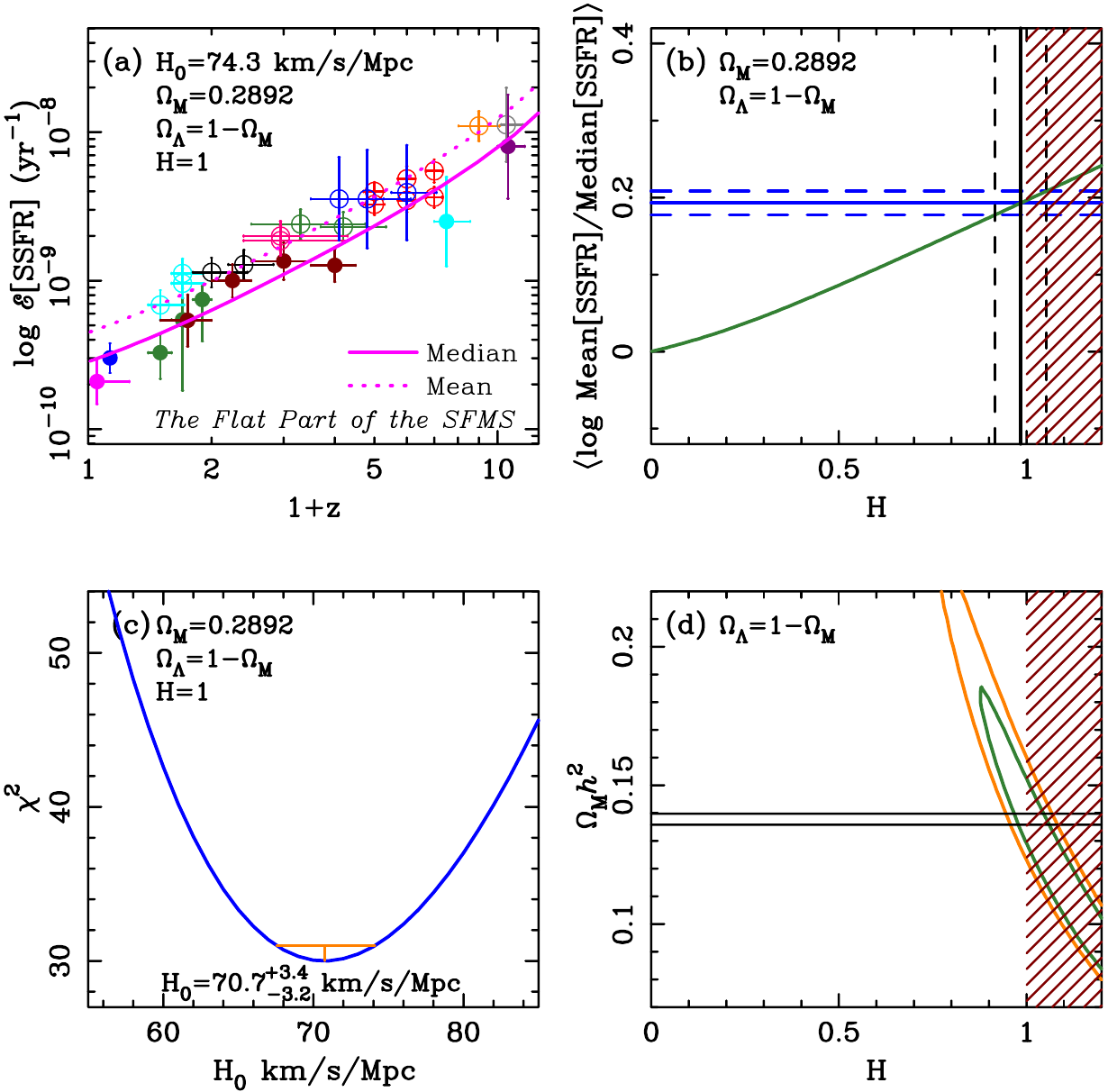}}
\caption{(a) Evolution of the median SSFR with redshift for lower-mass/bulgeless galaxies ($9.0\simlt{\log{M}}\simlt{10}$,
depending on the redshift). Filled circles show {\it median\/} SSFRs from the literature for deep, representative
samples of galaxies on the portion of the SFMS where there is no significant dependence of SSFR on stellar mass. Open
circles are similar measurements for {\it mean\/} SSFRs. We require sufficient depth in the samples to ensure that they
are derived from unbiased samples. Symbols are: Blue Filled: \cite{salim2007}, Violet Filled: \cite{bauer2013}, Green
Filled: \cite{karim2011}, Maroon Filled: \cite{kajisawa2010}, Cyan filled: \cite{mclure2011}, Grape Filled:
\cite{zheng2012}, Red open: \cite{gonzalez2014}, Orange open: \cite{labbe2013}, Green open: \cite{reddy2012}, Cyan open:
\cite{zheng2007}, Gray open: \cite{oesch2014}, Black Open: \cite{juneau2005}, Blue Open: \cite{lee2014}, Violet Open:
\cite{rodighiero2014}. Errors on the individual data points are a combination of the reported formal errors and
estimates of systematic uncertainties.
The evolution predicted for the median SSFR is shown by the violet solid line, assuming WMAP9 cosmological parameters
\cite{hinshaw2013}. The dotted line traces the expectation for the mean SSFR with redshift, where the difference between
the mean and median depends on the assumed $H$.
(b) Fitting the open and filled circles separately for their dependence on $1/T$ provides strong
constraints on $H$. The mean difference in $\log\text{SSFR}$ between the two samples is shown by the blue solid line,
with the standard error in the mean shown by the blue dashed lines. How this offset is expected to vary with $H$ is shown
by the green line. The red hatched region where $H>1$ is excluded because fBm diverges.
With these data we derive $H=0.98\pm 0.07$, shown by the vertical black lines, consistent with the
expectation of $H=1$.
(c) Assuming $H=1$ and WMAP9 cosmological parameters we fit the mean and median SSFR data for the Hubble constant,
finding $H_0=70.7_{-3.2}^{+3.4}$ km/s/Mpc with a systematic uncertainty of $5$ km/s/Mpc.
(d) Assuming a flat cosmology, the data can be fit to simultaneously constrain $H$ and $\Omega_Mh^2$. The
value of $\Omega_Mh^2=0.1378\pm 0.0020$ from WMAP9+SPT+ACT+SNLS3+BAO \citep{hinshaw2013} as shown by the horizontal
black lines. Contours at 1- and 2-$\sigma$ are shown by the green and orange lines respectively.
\label{fig:pull4}}
\end{figure*}

\section{Implications}
\label{sec:predict}

While $H$ may appear unconstrained, intuition makes $0.5\ll{H}\le{1}$ likely. The futures of galaxies are
changed permanently by all sorts of astrophysical processes, with star-formation ``remembering'' that
a galaxy, e.g., accreted gas-rich satellites, experienced a major merger, or fell into a larger halo.
Galaxies in higher density regions of the universe forever remain in such higher density regions. The steady average
increase in matter density over time brings galaxies from lower density regions to higher density, never to return.
Such reasoning suggests that $S$ should display something close to maximal persistence, with $H\cong 1$.
Interestingly, \cite{vergassola1994} derived a relationship between $H$ and $\alpha$, the faint-end slope of
the mass function, such that $\alpha=-H-1$ when exploring origin of the mass function. And the fact
that the deepest samples now reach $\alpha=-2$ \citep[e.g.][]{tomczak2014} also suggests such an extreme value for $H$.

We now explore some of the basic ramifications that arise from our derivations, including direct
constraints on $H$, surprisingly useful aspects of the predicted time-dependence of SFMS, and implications for the scatter
in SSFR.

\begin{figure*}[t]
\centerline{\includegraphics[width=7.0in]{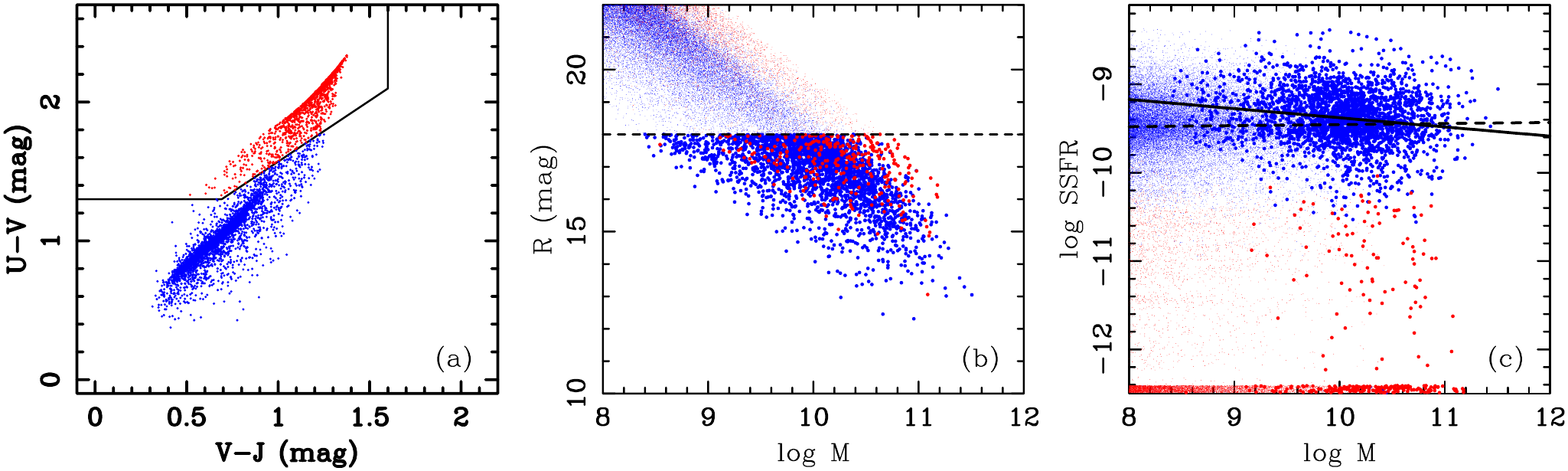}}
\caption{(a) $UVJ$ bicolor diagram for models of that begin forming stars at $z=20$, assuming $H=1$, no dust, and
solar metallicity \citep{bc2003}. Observational definitions for quiescent and star-forming galaxies are shown in black
\citep{williams2009}, and these are used to split the models into red and blue for further analysis.
(b) The correlation of predicted $R$-band magnitude with stellar mass, assuming the present day stellar masses
are drawn from a \cite{schechter1976} mass function with $\alpha=-1.4$ and $\log{M*}=10.85$, and the galaxies are distributed
randomly in distance over $0.02<z<0.2$. An selection limit at $R=18$ mag to mimic the SDSS, with smaller point symbols
used for galaxies fainter than this limit
(c) SSFR vs stellar mass, with the SSFRs of galaxies at $S=0$ clipped for illustrative purposes only.
Using all the points, a linear least-squares fit to all of the $UVJ$-selected
star-forming models yields the dashed line, with no apparent slope in $\log\text{SSFR}$ with stellar mass. But because
of the significant scatter in SSFR at fixed mass, the selection at $R=18$ mag imposes a bias on a simple linear least
squares fit, shown by the solid black line, with its slope of $-0.14$ dex/dex. The biweight estimate for the scatter
about this fit is 0.3 dex.
\label{fig:uvj}}
\end{figure*}

\subsection{$\langle{SSFR}\rangle$ and Its Evolution}
\label{sec:sfms}

Figure \ref{fig:pull4}(a) compiles median and mean SSFRs from the literature for galaxies in the portion of the SFMS
where $\langle\text{SSFR}\rangle$ is relatively independent of galaxy mass. We only show mean or median SSFRs derived
from galaxy samples with sufficient depth to derive unbiased measurements. In particular this restriction applies to the
original targeting of galaxies originally selected to be surveyed, and an additional criterion that SSFRs were measured
for their full samples, without regard to ongoing star-forming activity. Such data then represent all galaxies in the
given mass bin, sampling the full spans of low and high SSFRs. The violet solid line in Figure \ref{fig:pull4}(a) shows
Equation \ref{eq:clt4h}, adopting $H=1$, $z_\text{start}=20$, $\Omega_M=0.2892$
\citep[WMAP9+SPT+ACT+SNLS3+BAO;][]{hinshaw2013}, $\Omega_\Lambda=1-\Omega_M$, and a Hubble constant of $H_0=74.3$
km/s/Mpc \citep{freedman2012}. The dashed line shows the expectation for the mean SSFR over time.

Care must be taken to distinguish between the mean and median given the non-Gaussian (and non-lognormal)
distribution of SSFRs. For $H=1$, the mean SSFR is higher than the median by a factor of
\begin{eqnarray}
{\text{Mean}[S/M] \over \text{Median}[S/M]} = x_1 + x_2
\end{eqnarray}
\text{where}
\begin{eqnarray}
x_1&=&{1\over \sqrt{2\pi}}\biggl[e^{-{1\over 2}}-1\biggr] + {1\over 2}\biggl[
\text{erf}({0}) + \text{erf}\biggl({1\over\sqrt{2}}\biggr)\biggr]\\
x_2&=&{e^{1\over 2}\over 2}\biggl[1+\text{erf}\biggl({1\over \sqrt{2}}\biggr)\biggl]
\end{eqnarray}
This ratio is $\sim 1.57$, or $\sim 0.196$ dex. The dependence of this systematic logarithmic offset between mean and
median SSFR depends on $H$ in a manner shown by the green line in Figure \ref{fig:pull4}(b).

Note that $H$ is the only free parameter --- the value is not tuned to the observed
properties of the galaxies at all, though, as stated earlier, our intuition strongly favors $H\approx 1$.
SSFR is defined with respect to the mass in stars still alive, so in plotting the predictions
we correct $M$ to stellar masses using \cite{bc2003} and a \cite{chabrier2003} IMF. The apparent agreement between the
observed and predicted evolution for the median and mean SSFR appears surprisingly good, especially given
the absence of physics or arbitrary parameters for tuning. We now proceed to quantitative tests of the predictions.

The data constrain $H$ in two ways. Because median SSFR is predicted to evolve according to $\mathcal{E}[\text{SSFR}]\propto
(1+H)/\mathcal{T}$, normalizing the violet solid line to the data provides a joint constraint on $H$ and on $H_0$
(assuming $\Omega_M$ and $\Omega_\Lambda$).

But the data also constrain $H$ independent of $H_0$ because there are sufficient numbers of
independent measurements of both mean and median SSFRs over cosmic time. The difference between the mean and median is a
is a simple function of $H$ through the non-Gaussian distribution in SSFR. Therefore we
fit curves of $1/T$ to the mean and median SSFRs to calculate the mean logarithmic offset between the
mean and median SSFRS. The blue lines in Figure \ref{fig:pull4}(b) mark the measured offset of $0.193\pm 0.015$ dex
between the mean and median SSFRs. As stated above, the green line traces the dependence of this offset on $H$, so the
intersection of the blue and green lines at $H=0.98\pm 0.07$ provides the best-fit value for star-forming (and
bulgeless) galaxies over cosmic time. These results confirm that $H\cong{1}$ in a way that is independent of $H_0$. As
more observations of distant galaxies accumulate, constraints on $H$ should improve.

Assuming all histories of $S$ begin at the same time, $\mathcal{E}[\text{SSFR}]$ should be a standard clock with which
one can infer cosmological parameters.
In Figure \ref{fig:pull4}(c) we fix $\Omega_M$ and $\Omega_\Lambda$, fit the data, and find
$H_0{(1/2+H/2)}=70.7^{+3.4}_{-3.2}$ km/s/Mpc, with an estimated systematic uncertainty of $\sim 5$ km/s/Mpc.
As better data are accumulated, and greater care taken to quantify the intrinsic distributions of SSFR at fixed mass,
these uncertainties should improve. Improvements in both sample gathering and in their analysis should also improve the
systematic uncertainties in both SFRs and stellar masses. Observations over a
long range of redshifts, with multiple SFR indicators may mitigate systematic errors in SSFR. In fact, the constant
relative scatter in SSFR implies that formal errors in in $\mathcal{E}[\text{SSFR}]$, and thus in $H_0$, should scale as
$1/\sqrt{N}$.

Since the data are direct measures of lookback time with redshift, the cosmological constraint is largely on
$\Omega_Mh^2$. In
Figure \ref{fig:pull4}(d) we show $\chi^2$ contours in a plot of $\Omega_M h^2$ vs the Hurst parameter $H$.
For $H=1$, we derive $\Omega_M h^2=0.138_{-0.013}^{+0.014}$, consistent with the $\Omega_M h^2=0.1378\pm 0.0020$ derived
from WMAP9+SPT+ACT+SNLS3+BAO \citep{hinshaw2013}, shown by the horizontal black lines.
For larger values of $\Omega_Mh^2$ values of $H<1$ are statistically permitted given the
the relatively large uncertainties in the SSFR measurements. Improvements in
data quality and sample depth should translate directly into tighter constraints on these parameters.

Deriving a new cosmological constraint was not the intent of this paper, so
let us take a moment to step back from this rather detailed digression on cosmology, and reflect on
the key point. We began with one assumption: that the star-formation rate
of a galaxy is probably (probabilistically) constant from one brief epoch to the next. Events perturb these
growth rates, and the resulting stochastic changes to stellar mass growth
are correlated over long and short timescales.
By themselves these two rules constrain how the median SSFR of the ensemble of star-forming galaxies evolves with time,
and these rules also constrain the intrinsic scatter. Given the predicted scatter and distribution of SSFRs at fixed
mass, the difference between the median and mean SSFRs from the literature confirms $H=1$ within the errors. With great
accuracy Equation \ref{eq:clt4h} describes the evolution of the median (and mean) SSFR
for the ensemble of star-forming disks over time. By adopting $H=1$, one even finds
that the expectations for the mean/median SSFR can provide strong constraints on $H_0$ and $\Omega_M$, yielding results
in excellent agreement with canonical values.

Even with the data analyzed here, the formal and systematic errors on cosmological parameters are surprisingly small,
with 5\% random and 7\% systematic uncertainties in $H_0$.
In the future, greater care with such data may provide constraints that are competitive, but
extreme care must be taken to minimize selection biases. Selecting galaxies based on their on-going SFRs
will yield samples from the top half of the intrinsic scatter \citep[e.g.][]{elbaz2011}, potentially biasing $H_0$ low.
Such a selection may also reduce the observed scatter in SSFR, lulling the observer into thinking their
$\sigma/\sqrt{N}$ error bar was unreasonably small. Care should also be taken to avoid SFR indicators with sensitivity
to long timescales, as mentioned earlier. Star-formation indicators that are sensitive to $\sim 1$ Gyr timescales
convolve the stellar mass growth from $z=20$ through $z=5$, and from $z=5$ through $z=3$.

Both SFR and $M$ are sensitive to the initial mass function (IMF). But because they are equally dependent on
the assumed shape at the low-mass end, the form and shape of the violet line in Figure \ref{fig:pull4}(a) should
remain unaffected by potential bottom-heaviness IMFs in massive galaxies \citep[e.g.][]{conroy2013}.
Because the conversion of
an observed SF indicator to SFR does depend on the abundance of high-mass stars \citep{kennicutt1998a}, the tracing of
$\langle\text{SSFR}\rangle$ over time may constrain the universality and constancy of the high-mass slope.

\begin{figure*}[t]
\centerline{\includegraphics[width=7.0in]{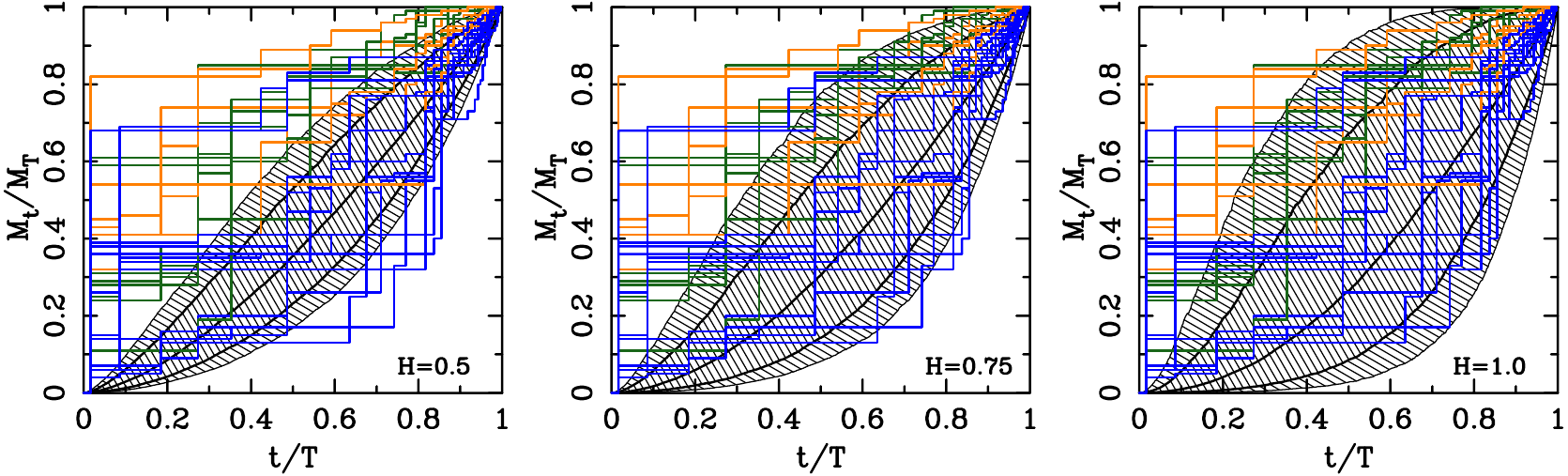}}
\caption{The 5th, 16th, 50th, 84th, and 95th percentiles of stellar mass relative to present-day are shown
for stochastic star-formation histories. The stellar mass growth derived from multiple HST pointings in nearby dIrr,
dTrans, and dE galaxies are overlaid using blue, orange, and green lines \citep[see][]{weisz2014}.
Even though the Local Group is probably not the most representative volume of the
universe, The distribution of dwarf galaxy histories is well matched by
fBm models with $H=1$, consistent with the value derived from the time evolution of $\langle\text{SSFR}\rangle$.
\label{fig:growth}}
\end{figure*}

\subsection{Dispersion in SSFR}
\label{sec:scatter}

Equation \ref{eq:clt5h} specifies the relative scatter in SFR at fixed mass. Using $H=1$,
\begin{eqnarray}
\text{Sig}[S/M]&\label{eq:sig}=&\mathcal{E}[S/M]\\
\text{Sig}[\ln{S/M}]&\label{eq:scatter}\approx &1\\
\text{Sig}[\log{S/M}]&\approx &0.434
\end{eqnarray}
Again, this scatter is defined with respect to the median SSFR.

Using galaxies in the Sloan Digital Sky Survey \citep[SDSS;][]{york2000} with coverage from the UV through the IR,
\cite{salim2007} derived an observed scatter of $\sim 0.5$ dex locally. Subtracting the mean measurement error of $\sim
0.2$ dex, one obtains $0.46$ dex as intrinsic.
The relative scatter is predicted by our derivations to remain constant with redshift, with, for example, estimates of
$\sim 0.5$ dex made out to $z=7$ \cite[e.g.][]{gonzalez2014}. Given the typical systematic and random uncertainties in
the measurements of SFR and stellar mass, this prediction of our model will remain difficult to verify with great
accuracy --- though data do indicate that the logarithmic scatter does not vary substantially over time and mass
\citep{salim2007,whitaker2012,gonzalez2011,stark2013}.

A number of authors appear to have constrained the scatter to smaller values, even as low as 0.2 dex
\citep[e.g.][]{brinchmann2004}. Some have achieved such remarkably low values by fitting out the systematic differences
between multiple indicators \citep{speagle2014}, with most work only focused on galaxies preselected as star-forming down
to modest limits in SFR \cite[e.g.][]{feulner2005}. Care must also be taken to avoid preselecting star-forming galaxies
\cite[e.g.][]{elbaz2011,guo2013}, as such a selection explicitly biases the distribution against galaxies with low SSFR,
or worse, biases the observations against low-mass galaxies with modest SFRs. That the relative scatter remains somewhat
constant with time and mass, not varying by orders of magnitude, is remarkable and speaks to the presence of an
underlying fundamental law. And while we have not ruled out the presence of astrophysical laws, we have reproduced
a number of the properties of the SFMS adopting only the central limit theorem as that law.

Particularly strong observational constraints on the scatter are difficult to obtain due to the
influence of selection effects. Figure \ref{fig:uvj} illustrates this difficulty by constructing a small simulation
of the SDSS volume. We use a random realization of 6400 fBm histories with $H=1$, assume $z_\text{start}=20$, solar
metallicity, and no dust attenuation \citep{bc2003}. Stellar masses were drawn from a \cite{schechter1976} mass function
with $\alpha=-1.4$ and $\log{M*}=10.85$, and then assumed to fill the volume between $0.02<z<0.2$ uniformly.

In Figure \ref{fig:uvj}(a), we show the resulting bicolor diagram of $U-V$ vs $V-J$ with the empirical separation into
quiescent and star-forming populations \citep{williams2009} shown in red (20\%) and blue (80\%). The correlation between
predicted $R$-band magnitude and stellar mass is shown in Figure \ref{fig:uvj}(b), along with a SDSS-like selection cut at
$R=18$ mag. Models fainter than this limit are shown using small symbols. Recall that this model does not (yet) include
bulge stars, or any {\it ex situ\/} stellar mass growth, which would serve to systematically depress SSFR at fixed mass.

Sample selection couples with the intrinsic dispersion to bias the observed scatter in SSFR. In this simulation
the RMS scatter in the full sample is 0.39 dex. For the subsample with $R<18$ mag the RMS scatter is 0.35 dex, while the
biweight estimate is $0.30$ dex. For studies at any redshift, the detailed selection criteria are as
critical as the method by which star-forming galaxies might have been selected.

The sample selection can also impose an apparent dependence of $\langle\text{SSFR}\rangle$ on stellar mass.
The selection at $R=18$ mag corresponds to $SFR\approx 0.1 M_\odot$/yr. This effective cut in SFR is made more clear in
Figure \ref{fig:uvj}(c), where SSFR is plotted against stellar mass. The black dashed line, with a slope of zero,
represents the best fit linear relation to the star-forming models regardless of $R$-band magnitude. But fitting to
the subsample with $R<18$ mag yields a slope of $-0.14$ dex/dex, shown by the solid black line. Note
that \cite{salim2007} found that lower-mass galaxies followed $\log\text{SSFR}\propto -0.17\log M$, and that
upon correcting SSFRs for bulge mass fractions \cite{abramson2014} also found a similar residual correlation
for star-forming disks. While we have not modeled the galaxy distribution and selection function of SDSS in great
detail, these simulations indicate that modest anticorrelations between $\langle\text{SSFR}\rangle$ and stellar mass
may very well arise from the interplay between sample selection and the dispersion in SSFR.

As astronomers, we may expect that the scatter in $S/M$ should be affected by long-term trends in bulge formation,
feedback, environmental processes, or any correlation of $z_\text{start}$ with galaxy/halo mass. Indeed these will all
produce systematic deviations over time in $\mathcal{E}[S/M]$ from the expectation values given above. But because the
central limit theorems still hold for stochastic processes centered on long-term expectations (such as the long-term
imposition of nonnegativity on $S_t$), our formalism need merely be superimposed on long-term trends when (or where)
such particular astrophysical processes matter. In other words, the second term in Equation \ref{eq:var2} becomes
operable, but the relative scatter, defined as $\text{Sig}[S/M]/\mathcal{E}[S/M]$, remains unaffected so long as one's
samples remain fair and representative of star-forming (or potentially star-forming) galaxies in the survey volume.

Some implications of such dispersion for the evolution of the stellar mass function \citep[e.g.][]{munoz2014}
are investigated below, with more in-depth analysis saved for later \citep{kelson2014b} in the context of the
Carnegie-Spitzer-IMACS redshift survey \citep{kelson2014a}.

\subsection{The Stochastic Histories of Dwarf Galaxies}
\label{sec:sfhs}

The star-formation histories of
nearby dwarfs serve as useful tests of our predictions. Figure \ref{fig:growth} overlays recently published SFHs for
dIrr, dTrans, and dE galaxies from \cite{weisz2014} on the distributions of stochastic models with varying
values of $H$. We excluded the dSph galaxies in their sample; these truncated early, in ways not captured by
assuming stochasticity through today.

Qualitatively the model distributions provide a good match to the broad distribution of star-formation histories for
dwarf galaxies. The percentiles for $H=1$ encompass the dwarf histories better than $H=0.5$ or $H=0.75$, so we take
these results as supportive of the modeling. In Figure 14 of \cite{weisz2014}, the models of \cite{behroozi2013} compare
far less favorably.

These models have been derived with star-forming/disk galaxies in mind --- galaxies that have not accreted appreciable
amounts of stellar mass formed {\it ex situ\/}. We will explore some simple extensions to the formalism to encompass
{\it ex situ\/} mass growth in later sections but, for the moment, we proceed to a comparison of the model histories
to that of the nearest disk galaxy.

\subsection{The Milky Way}
\label{sec:mw}

The Milky Way galaxy provides a unique window into the star-formation history of a fairly massive disk galaxy, given
that the bulge, if there is one, is a small component of the mass and likely formed secularly \citep{shen2010}.
With a current rate of star-formation of $\sim 2$ $M_\odot$/yr and a stellar mass of $\sim 5\times 10^{10} M_\odot$,
$\log\text{SSFR}_\text{MW}=-10.4$ dex \citep{snaith2014}. If the median SSFR for galactic disks is
$\log\langle\text{SSFR}\rangle=-9.55$ at $z=0$ ($H=1$), then the Milky
Way's SFR is a factor of 7 too low compared to star-forming disks ($\sim 2\sigma$; Equation \ref{eq:scatter}).

Other diagnositcs, such as the Milky Way's angular momentum, also point to the Milky Way as having a relatively
boring late-time history compared to other comparable spirals \citep{hammer2007}. Using the broken power-law form of the SFMS
from \cite{salim2007}, one infers that star-forming galaxies with the mass of the Milky Way have an average
$\langle\text{SSFR}\rangle=-10.25$. But galaxies with the mass of the Milky Way have a median bulge mass fraction of
40\% \citep{abramson2014}, such that disks comparable to the Milky Way have
$\langle\text{SFR}/M_\text{disk}\rangle=-10.0$, and implying that the Galaxy is only
a factor of 2-3 low in relation to comparable systems.

\begin{figure}[h]
\centerline{\includegraphics[width=2.9in]{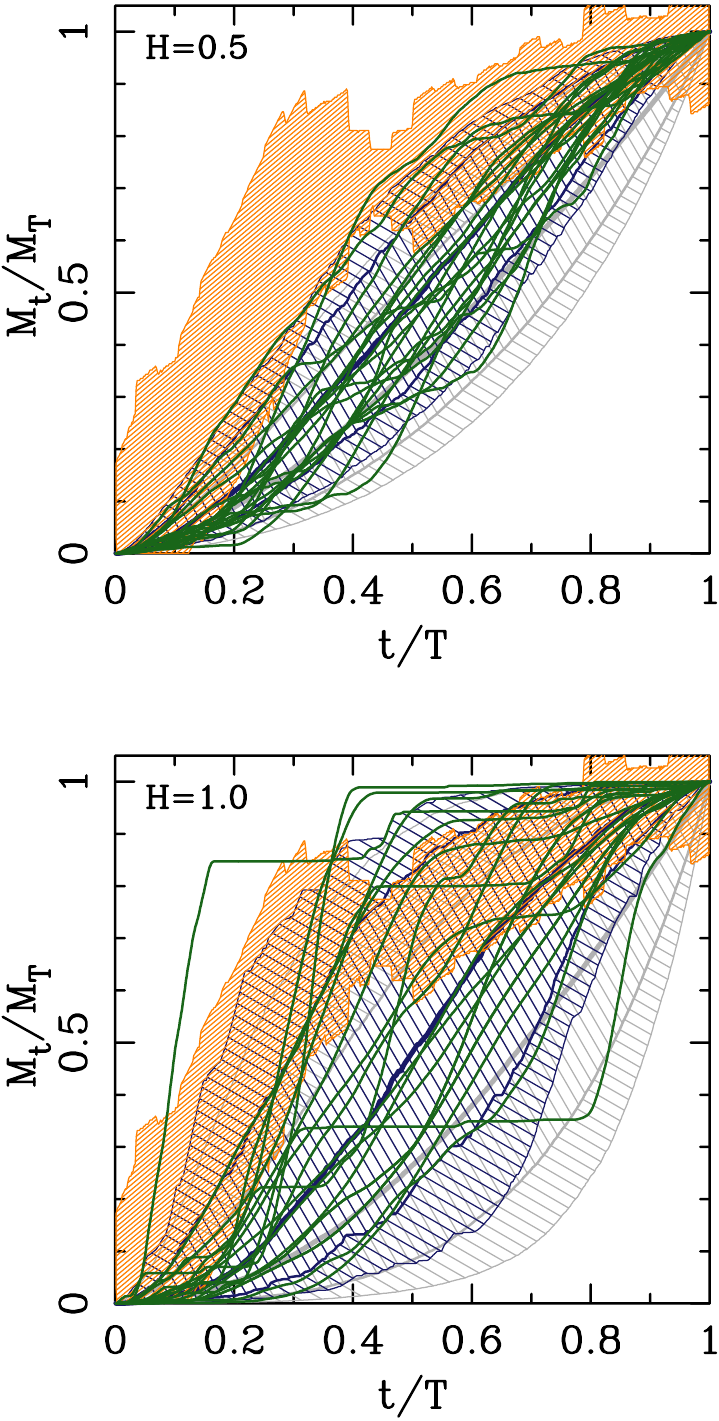}}
\caption{The growth of the Milky Way (orange) \citep[data taken from][]{snaith2014}, overlayed on the distributions
of model growth histories in gray for $H=0.5$ (top) and $H=1$ (bottom). The dark blue show
the distributions of those disk galaxies that would have specific star-formation rates within 50\% of the Milky Way's
SSFR at $t=T$ (where $T=$ today). Twenty random histories from this distribution are shown in green, to illustrate any
similarities between the diverse fBm SFHs of galaxies with similar specific star-formation rates to the Milky Way today.
Relative inactivity at the present-epoch restricts the range of SFRs that the Galaxy likely had in the past when $H=1$
due to long-term covariance between stochastic changes in SFR. The suite of growth histories when $H=1$ suggests that
Milky Way-type SFHs are not particularly rare.
\label{fig:snaith}}
\end{figure}

In Figure \ref{fig:snaith} we overlay the mass growth of the Milky Way from \cite{snaith2014} on the distributions
of mass growth for $H=0.5$ and $H=1$. The full distributions of 
model growth histories are shown using light gray. In dark blue we restrict these
distributions to those that have SSFRs within 50\% of the Milky Way's at the present epoch. The difference between
the Brownian case of $H=0.5$ and the fBm case with $H=1$ is striking, with Milky Way-type histories virtually
non-existent when there is no long-term correlation between stochastic events. But using the $H=1$ models, which also
best matched the evolution of $\langle\text{SSFR}\rangle$ with time (\S \ref{sec:sfms}), we see that the Milky Way did
not experience a star-formation history particularly out of the ordinary. When $H=1$, $\sim 9\%$ of the fBm histories
have SSFRs within 50\% of the Milky Way's, and, interestingly, \cite{hammer2007} suggested that only $\sim 7\%$ of
spiral galaxies were like the MW given its angular momentum and stellar mass.

From this qualitative comparison, we would conclude that, while $\sim 90\%$ of disk galaxies have SSFRs higher than
the Milky Way, its star-formation history is broadly represented by those stochastic models with comparably low SSFR
at late times. About $1/3$ of these model histories ``experienced'' rapid growth at early times, but
long-term covariance between stochastic changes in SFR left diminished rates of {\it in situ\/} stellar mass growth at
late times. The questions, also invoked by \citep[][]{snaith2014}, then become: What event, or events, occurred 8 Gyr ago
to stochastically shutdown star-formation in the Milky Way? What interactions may have occurred when the Local Group
first formed \citep{forero2011}? And did this new environment serve to stifle the fueling of stellar mass growth for a
Gyr, leaving it depressed long thereafter? In \S \ref{sec:time2} we discuss some of these questions
when revisiting the issue of variable timescales of stochasticity.

The consistency between the Milky Way's star-formation history and the models, the agreement between the observed
and predicted evolution of $\langle\text{SSFR}\rangle$ over cosmic time, and the agreement between the
observed scatter in SSFR and that predicted by our derivations, all boost our confidence in this statistical approach to
the evolution of disk galaxies. Our derivations specifically led to an intrinsic distribution in specific star-formation
rates and SFHs at fixed mass, and these have direct consequences for the distribution of stellar $M/L$ ratios in disk
galaxies. After exploring the effects of stochasticity on the $M/L$ ratios of galaxies,
we will discuss the evolution of the stellar mass function and ways of extending the formalism
to the broader distribution of quiescent and star-forming systems.

\subsection{The Scatter in the Tully-Fisher Relation}
\label{sec:ml}

\begin{figure}[ht]
\centerline{\includegraphics[width=3.0in]{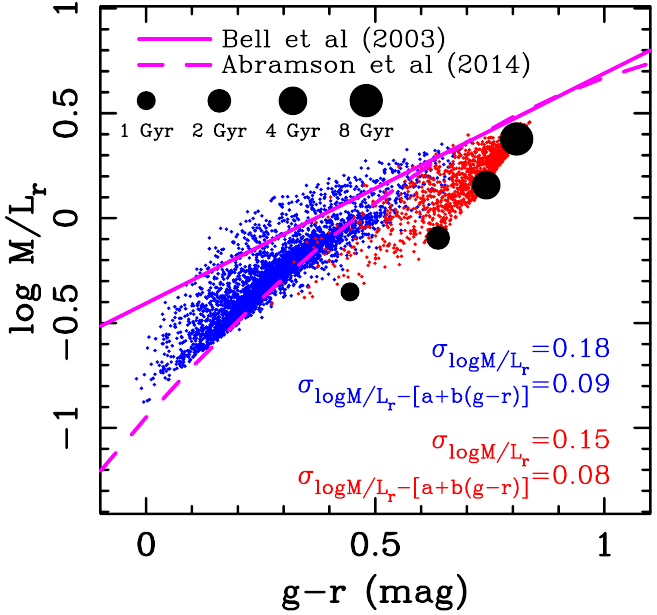}}
\caption{(a) For the same models as in Figure \ref{fig:uvj} we plot predicted $r$-band $M/L$ ratios vs stellar mass.
The biweight estimate of the scatter in $\log M/L_r$ is 0.18 dex and 0.15 dex for the star-forming and quiescent
stellar disks, respectively. (b) The $r$-band $M/L$ ratios of the models are plotted against $g-r$ color, showing that
these models also reproduce the correlation derived by \cite{bell2003} for star-forming galaxies (violet).
The scatter in the models about the \cite{bell2003} correlation is 0.09 dex and 0.08 dex for star-forming and quiescent
disk galaxies.
\label{fig:mlplot}}
\end{figure}

The broad diversity of SFHs for disk galaxies has implications for the Tully-Fisher relation \citep{tully1977}
and its intrinsic scatter. Figure \ref{fig:mlplot} takes the distribution of model galaxies
from Figure \ref{fig:uvj} and explores their $M/L$
ratios as a function of color. Recall that these models were generated with no variation in metallicity and no dust
attenuation. Before discussing the detailed scatter in $M/L$ ratios we overlay the derived correlation of $M/L$ with
galaxy color from \cite{bell2003} in violet, adjusting to the \cite{chabrier2003} IMF. Using the DR7 SDSS
data on star-forming galaxies \citep{brinchmann2004}, \cite{abramson2014} attempted to remove the bulge mass
from the global $M/L$ ratios and colors and derived a correlation between $M/L$ and $g-r$
specifically for the underlying disk mass, shown by the violet dashed line.

The agreement between our simple models and the trend from \cite{abramson2014} is excellent for actively star-forming
systems. The comparison with \cite{bell2003} is less favorable though they remarked that recent bursts of star-formation
would serve to move their locus to lower $M/L$ ratios, bringing their correlation with color in better agreement with
ours. Overall, however, the trends are similar, indicating that if one implemented our SFHs to derive stellar mass
Tully-Fisher relations, one would recover stellar mass Tully-Fisher relations that are qualitatively similar to those
previously published \citep{bell2003,mcgaugh2005}.

More importantly, however, is the retrodiction for the scatter in $M/L$ ratios given the diversity in SFHs and
dispersion in on-going SSFRs. The SFHs of disk galaxies that would be classified as actively star-forming today (blue)
have a scatter of 0.18 dex in $\log M/L_r$, equivalent to 0.45 mag in an $r$-band Tully-Fisher relation. Those models
bluer than $g-r<0.5$ have a scatter equivalent to 0.40 mag. These estimates are in good agreement with the 0.42 mag
intrinsic scatter estimated by, e.g., \cite{pizagno2007} --- though systematic correlations of bulge mass fractions,
metallicity, and dust content with galaxy mass may serve to reduce the scatter, as would selection biases.

The low predicted scatter of 0.15 dex for quiescent disks is also important. For reference, we have marked
positions in Figure \ref{fig:mlplot} for single stellar populations (SSP) with ages of 1, 2, 4, and 8 Gyr. Using
only those quiescent disks redder than the 2 Gyr mark, one finds a scatter in $\log M/L_r$ of 0.10 dex. For those
redder than the 4 Gyr mark, the scatter is $0.06$ dex. Thus our models for the SFHs of star-forming disk galaxies
have consequences for the fundamental plane of early-type galaxies \citep{dress7s,dd87}. The fundamental plane is an
observed correlation between $M/L$ ratio and mass (and size) for elliptical and S0 galaxies, and the intrinsic scatter
in $M/L$ is small, $0.06-0.10$ dex in restframe optical passbands such as $V$ or $r$
\citep{jfk96,kelson2000,labarbera2010}. Such a low level of scatter has been used to infer that the scatter in
luminosity-weighted ages for the stellar populations of early-types must be quite small, $\ll 20\%$
\citep[e.g.][and many others]{kelson2000}.

As one can see from these models, disks that stochastically stop forming stars for an appreciable amount of time
fade and redden swiftly. The result is that the relative uniformity of $M/L$ ratios is preserved, even for
quiescent disk galaxies --- thus scaling relations like the Tully-Fisher relation or the fundamental
plane for quiescent galaxies should be observable at nearly any epoch \citep{miller2012,toft2012}.
Furthermore, old galaxies that are stochastically rejuvenated will not fall far from the fundamental plane for
very long. For example, an 8 Gyr old SSP that has a 5\% (by mass) infusion of new stars will only be 0.1 dex in $M/L$
ratio from the $r$-band fundamental plane after 1 Gyr. After 2 Gyr this remnant would be 0.05 dex away from where it
used to sit in the fundamental plane.

To summarize: rejuvenating old galaxies stochastically with new stellar mass will move galaxies off the fundamental
plane, but only temporarily \cite[e.g.][]{kelson1997}. During that time, the galaxy would have a stellar mass-to-light
ratio in line with the Tully Fisher relation. If that star formation continues, the galaxy would
simply continue to grow, and maintain normal, ordinary $M/L$ ratios for disk galaxies. If that star-formation ceases
through normal stochastic processes, then it will happily retire for half a Hubble time (see below) to the seemingly
calmer pastures of the fundamental plane. Such a scenario would naturally lead to stellar populations being
more dependent on galaxy mass, or velocity dispersion, than on galaxy morphology itself \cite[e.g.][]{robaina2012}.

\begin{figure*}[t]
\centerline{\includegraphics[width=5.0in]{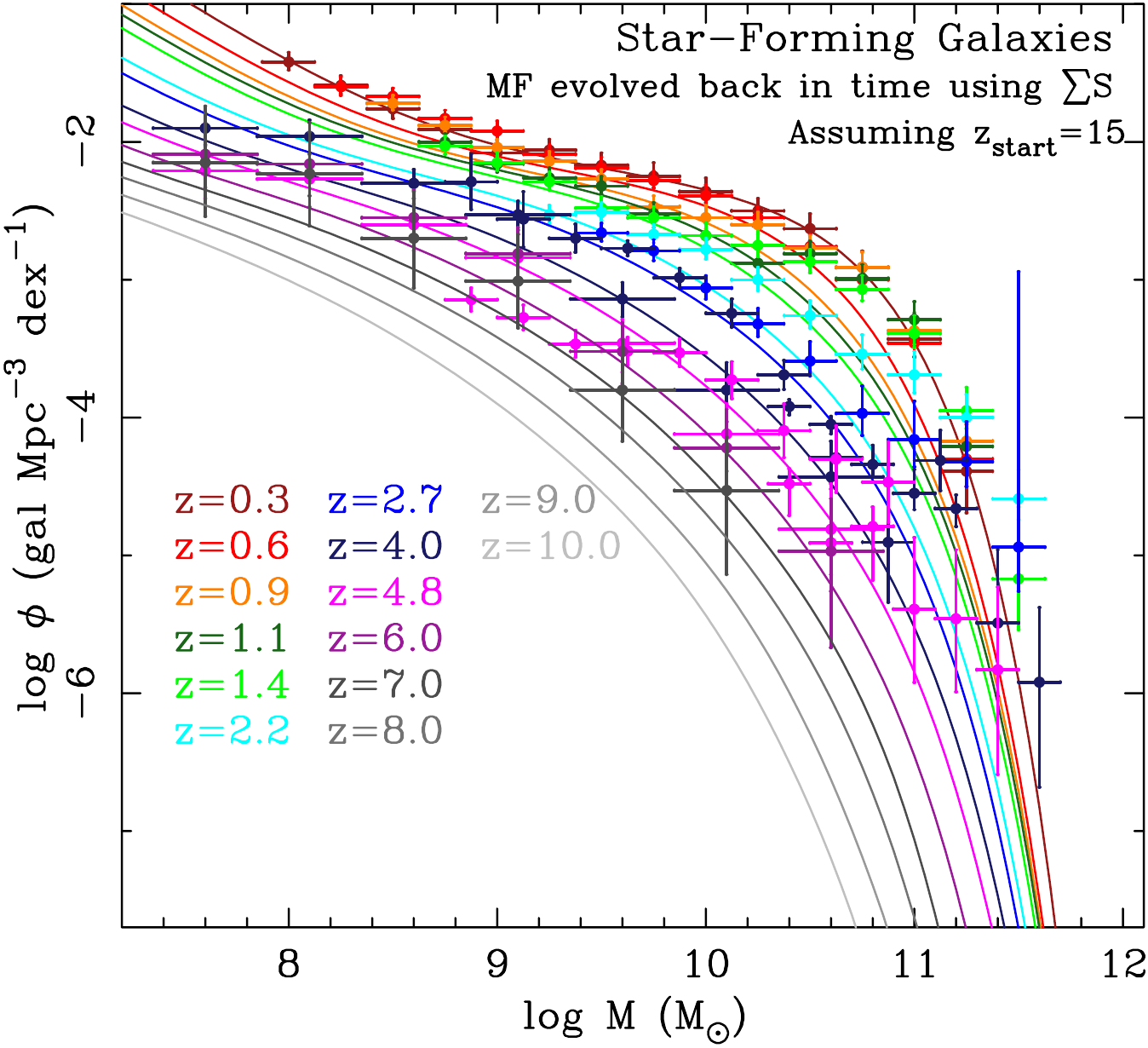}}
\caption{Measurements of the stellar mass function of star-forming galaxies back to $z\sim 7$
\citep{tomczak2014,lee2012,gonzalez2011,caputi2011}. Starting with the double \cite{schechter1976} function from
Z-FOURGE at $0.2<z<0.5$ \citep{tomczak2014}, we use the fBm realizations to evolve the mass function back another 10 Gyr.
While the match is not perfect, notice the transition from a single \cite{schechter1976} function form at early times
to the more pronounced knee at late times. This bunching up of stellar mass at $M*$ arises because $\mathcal{E}[M]$
evolves more slowly than one would naively assume if all galaxies had the same specific star-formation rate. No
quenching was added to these expectations outside of that expected for SFHs with significant long-term correlations
between stochastic changes in SFR. The mass functions at $z>7$ are very sensitive to the assumed starting redshift,
where the ones shown here are derived using $z_\text{start}=15$.
\label{fig:mfevol}}
\end{figure*}

\section{On the Evolution of the Stellar Mass Function}
\label{sec:mf}

The stellar mass function of star-forming galaxies is well described by a \cite{schechter1976} function, with a
power-law shape at low galaxy masses up to an exponential cutoff at masses above a
characteristic mass. Modern fitting of mass functions, for surveys that reach to low mass galaxies,
now include two \cite{schechter1976} functions \citep[e.g.][]{baldry2012,tomczak2014}. Such functional forms have been
used to characterize galaxy populations in the SDSS $z\sim 0$ \citep{moustakas2013} up through $z\sim 7$ in the HUDF
\citep{gonzalez2011}. Over the past 13 Gyr, the slope, $\alpha$, of the low-mass end of the mass function has stayed
remarkably constant, at $\alpha\sim -1.5$ \citep{gonzalez2011,tomczak2014}.

A uniform SSFR is often cited as a requirement for maintaining a constant $\alpha$ over cosmic time
\cite[e.g.][]{peng2010}. When all galaxies have the same SSFR, the constant $d\ln M/dt$ simply translates the
stellar mass function uniformly to higher logarithmic mass bins. But we have demonstrated that underlying the narrow
observed range of SFR at fixed stellar mass is a distribution of SSFRs. Furthermore, galaxies in this distribution may
be experiencing changes in SFR on a range of timescales. As a result, the evolution of the stellar mass function must be
revisited.

\subsection{The Evolution of the Stellar Mass Function for Star-Forming Galaxies}
\label{sec:sfmf}

In Figure \ref{fig:mfevol}(bottom) we show stellar mass function measurements back to $z\sim 7$
\citep{tomczak2014,lee2012,gonzalez2011,caputi2011}. Here we take the separation into quiescent and star-forming from
the respective articles, but assume that at $z>3$ the published total mass functions largely represent star-forming
galaxies.
Adopting the double \cite{schechter1976} function fit to the $0.2<z<
0.5$ data \citep[from Z-FOURGE;][]{tomczak2014}, we use the fBm models ($H=1$) to evolve the star-forming mass function back
in time. There is a noticeable mismatch at high Mass, presumably due to an absence of (1) bulge assembly, (2) mergers ({\it
ex situ\/} stellar mass growth), (3) mechanisms to restrict fuel supplies in massive galaxies/halos, (4) systematic
increases in the timescales for stochastic change in massive galaxies/halos, or (5) distinguishing between centrals and
satellites. Some of these issues are discussed in \S \ref{sec:allmf} when exploring how to extend the derivations for all
galaxies.

Despite these shortcomings, a number of salient features in the model evolution already appear. When evolved back in
time, the late-time double \cite{schechter1976} function naturally transitions to a single \cite{schechter1976}
function. This transition in the mass function shape occurs because the mass growth (Equation \ref{eq:mth}) is not
exponential. In models that assume a constant SSFR for galaxies, the mass function simply shifts to higher masses evenly
at all logarithmic masses. But the constant logarithmic bins in mass encompass ever larger and larger spans of stellar
mass at high mass, and the predicted evolution in stellar mass simply cannot keep up with the growth in the size of the
bins. Thinking about this evolution from early times to the present, galaxies piled up in the high mass bins, enhancing
the knee of the mass function --- simply because growth was not exponential.

If the disks of galaxies growth through stochastic star-formation histories, how might the deficiencies in the model (at
high mass) be repaired, such that specific astrophysical mechanisms be constrained? In the next several sections
we explore ways of extending the formalism.

\begin{figure*}[t]
\centerline{\includegraphics[width=5.0in]{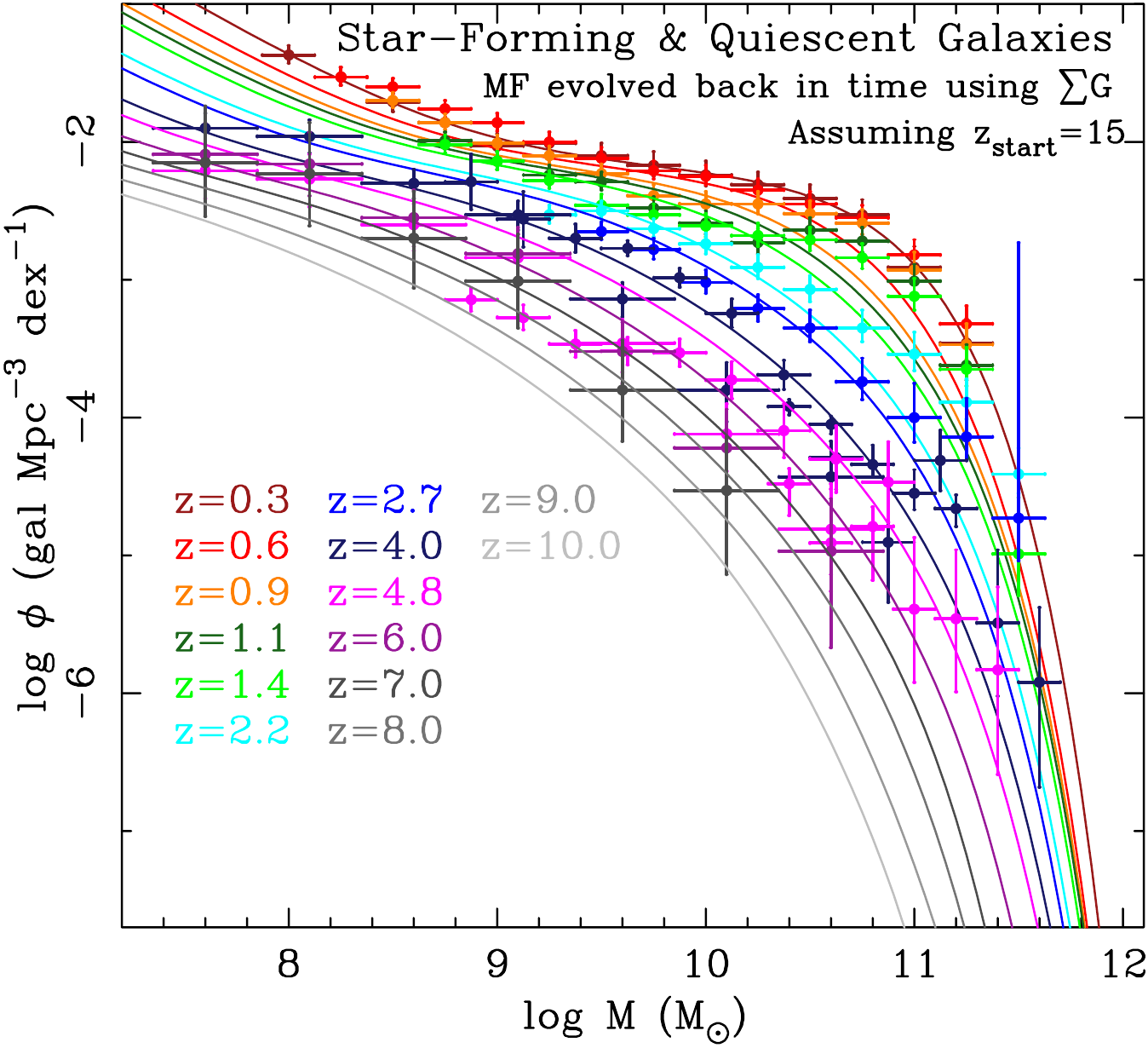}}
\caption{Same as in Figure \ref{fig:mfevol}(bottom) but for all galaxies, star-forming and quiescent.
\label{fig:mfevol2}}
\end{figure*}

\subsection{Towards an Evolving Stellar Mass Function for All Galaxies}
\label{sec:allmf}

Everything derived to this point has been based on the one assumption that star-formation occurs in disk galaxies as a
stochastic process with long-term correlations between stochastic changes in SFR. In this section we speculate on simple
extensions to our formalism for including {\it ex situ\/} stellar mass growth, of the kind that may assemble
galaxy bulges and spheroids. Essentially such mechanisms add only to $M$, and add nothing to $S$, thus depressing SSFR.
Only the evolution of the mass function can constrain these processes, as the SFMS is clearly degenerate between $S$ and
$M$. We include a discussion of the mathematical impact on the SFMS such that the effects of long-term systematic
reductions in fuel supply, or systematic changes in stochastic timescales can also be incorporated into the models in a
similar fashion.

\subsubsection{Writing All Mass Growth as a Stochastic Process}

Consider $G_{n,t}$, the amount of stellar mass added to galaxy $n$ at time $t$,
including both star-formation on site, and stellar mass created elsewhere but acquired through
merging or accretion. If $G$ is a stationary process, then
the earlier derivations already tell us:
\begin{eqnarray}
G_{n,T}&=&\sum_{t=1}^{T}(G_{n,t}-G_{n,t-1})\\
&=&\sum_{t=1}^{T}X'_{n,t}
\end{eqnarray}
and
\begin{eqnarray}
M_{n,T+1}&=\label{eq:all1}&\sum_{t=1}^TG_{n,t}\\
&=&\sum_{t=1}^T\sum_{i=1}^{t}X'_{n,i}
\end{eqnarray}
with the end result being the recovery of Equation \ref{eq:mth} (sadly we are never able to measure $G/M$).

Figure \ref{fig:mfevol2} takes the stellar mass function for all galaxies, star-forming and quiescent, from
\cite{tomczak2014} and evolves it back in time using the above general form, assuming all processes that contribute to
the growth of galaxies can be written as stationary stochastic processes. Thus instead of adopting the earlier
derivations for star-formation only, the models are assumed to describe the general mass growth histories of galaxies.
So long as all substantial aspects of stellar mass growth can be treated as stochastic processes their sum follows the
same derivations provided earlier (the sum of two martingales is also a martingale).

Thus Equation \ref{eq:mth}
can be applied to all galaxies --- so long as the stochastic changes are all stationary. It is this requirement that
simply must break to properly match the evolution in the mass function for galaxies at high masses at late times, or to
recover the change in slope of the SFMS at high masses \cite[e.g.][]{salim2007,whitaker2012,sobral2014}.

Qualitatively there are a number of features in the data that are reproduced by the model curves, but the process(es)
that shape the stellar mass growth of galaxies appear not to be constant over time or mass, as evidenced by the
late-time evolution of the SFMS at high-masses \citep{kajisawa2010,karim2011,whitaker2012,sobral2014}, and by the
late-time formation of galaxy groups and clusters \citep{williams2012}, in conjunction with the morphology-density
relation \citep[e.g.][]{dressler1980,vdw2007} or star-formation-density relation \cite[e.g.][]{patel2009,quadri2012}.

\subsubsection{Moving beyond a Single Stochastic Process}

So long as the total of the {\it in situ\/} and {\it ex situ\/} stellar mass growth for galaxies is a stationary
stochastic process, all of the physics of mass growth was subsumed into an unknown spectrum of $\sigma_{n,t}$. In order to
explore how astrophysical processes modify the SFMS and the mass function overt time, let us write the total accumulated
stellar mass growth by $T+1$ as
\begin{eqnarray}
M_{n,T+1}&=\label{eq:so}&\sum_{t=1}^TS_{n,t} + \sum_{t=1}^TO_{n,t}
\end{eqnarray}
Again $S_{n,t}$ is the stellar mass growth due to star-formation, but now $O_t$ is an additional component of growth
arising from outside, i.e. {\it ex situ\/} sources of stellar mass.

This representation is not equivalent to decoupling star-formation from mergers. Both mergers and satellite accretion
bring fuel for new {\it in situ\/} star-formation --- star-formation included in the SFMS through measurements of SFRs.
Such events also add old stellar mass directly to the
denominator of SSFR, with no direct way to account for it in diagrams of (S)SFR vs stellar mass. For example, when
mergers occur between two galaxies with gas mass fractions of 50\%, the old stellar mass being brought in is only 25\%
of the remnant, and as the original gas reservoir and the newly acquired gas are all fruitfully turned into new stars, the
old stellar mass acquired through the merger may be a small perturbation to the mathematics. Eventually this
accumulation depresses the observed SSFRs, and in ways that are not accounted for by the integration of SFR over cosmic
time.

But this example does not fully convey profound ambiguities that lie hiding in the mathematics.
If the ratio of gas to stars in galaxies is reasonably constant, or at
least there is a distribution of gas fractions that is stationary with time, about a mean $f$ then interesting things
happen with Equation \ref{eq:so}. Let us rewrite it using star-formation efficiency and fuel supply:
\begin{eqnarray}
M_{n,T+1}&=\label{eq:ffp}&\sum_{t=1}^T\epsilon_{n,t}[F_{n,t} + F'_{n,t}] + P_{n,t}
\end{eqnarray}
We have split $O$ into a portion that is newly acquired gas, and the mass in stars being acquired; $F_{n,t}$ is the gas
supply already on hand, $F'_{n,t}$ is the new gas being brought in, and $P_{n,t}$ is the mass in stars being accreted
along with the new gas. Equation \ref{eq:ffp} can now be written as
\begin{eqnarray}
M_{n,T+1}&=\label{eq:fff}&\sum_{t=1}^T\epsilon_{n,t}[F_{n,t} + F'_{n,t}] + {(1-f)\over f}F'_{n,t}\\
&=\label{eq:ff}&\sum_{t=1}^T\epsilon_{n,t}F_{n,t} + \epsilon'_{n,t}F'_{n,t}
\end{eqnarray}
where we construct $\epsilon'_{n,t}$ as a modified efficiency.

The sum of these two stochastic processes is still a stochastic process, so Equation \ref{eq:ff}
is functionally identical to Equation \ref{eq:so}. And so long as gas fractions are themselves stationary stochastic
processes, the sum of the {\it in situ\/} and {\it ex situ\/} stochastic processes is itself a stochastic process,
covered by Equation \ref{eq:all1}, and equivalent to the forms derived in \S \ref{sec:process} and \S \ref{sec:fBm}.
When (a portion of) such stochastic differences are no longer randomly distributed about zero, such that
(sub)populations experience biased long-term stochastic changes
to stellar mass growth, can the aggregate process of galaxy growth no longer be reduced so simply.

As gas fractions decline with time \citep{tacconi2010,tacconi2013}, Equation \ref{eq:so} cannot be reduced to a
single stochastic process. Astrophysically, the fraction of old stellar mass accumulated through mergers and accretion may
be significant compared to the stellar mass created out of old and new gas, though the available gas supply may
also be declining over cosmic time. Assuming $H=1$ for both processes, which may not necessarily be
valid, Equations \ref{eq:sth}, \ref{eq:mth}, and \ref{eq:so} tell us that
\begin{eqnarray}
\mathcal{E}[S_T]&\label{eq:soth}=&\sigma_S\biggl({2\over{\pi}}\biggr)^{1/2}\biggl({T\over{2}}\biggr)\\
\mathcal{E}[M_T]&\label{eq:moth}=&\sigma_S\biggl({2\over{\pi}}\biggr)^{1/2}\biggl({T^2\over{4}}\biggr)
 + \sigma_O\biggl({2\over{\pi}}\biggr)^{1/2}\biggl({T^2\over{4}}\biggr)\\
\mathcal{E}[S_T/M_T]\label{eq:smoth}&=&2\over{T(1+\sigma_O/\sigma_S)}
\end{eqnarray}
for galaxies with the variances in the stochastic changes in $S$ and $O$ described by 
$\sigma^2_{S,n,T}$ and $\sigma^2_{O,n,T}$.

Of course the stars that make up $O_{n,t}$ were made {\it somewhere\/}. How much of that star-formation was measured
as {\it in situ\/} stellar mass growth in other galaxies at earlier epochs, and in galaxies with what masses? In
evolving the mass function back in time, how does one redistribute the modeled amount of accreted stellar mass back
into the stellar mass function at earlier epochs? The SFMS and stellar mass functions do not provide sufficient
information but additional data such as the evolution of the SFRD \citep{madau1996,lilly1996,cucciati2012,bouwens2014} and
galaxy clustering as a function of mass \citep{quadri2007,coil2008} will be crucial.

Properly accounting for merging in these scenarios may be difficult but there is a burgeoning area of mathematics that
may help in the future, that of subfractional Brownian motions \citep{bojdecki2004}. These are further generalizations
for particle systems where processes may not be stationary, and where branching occurs \citep{bojdecki2012}, which, is
simply merging with time reversed. Reversing time in such models is not as complicated as it might seem since reverse
martingales also obey the central limit theorem \citep{hall1980}. Exploring these formalisms is clearly beyond the scope
of this paper, but such new mathematical tools may prove to be useful over the next several years.

Additionally there is a natural inclination to assign $S$ to the disk, and $O$ to the bulge, and therefore assume one
can derive distributions of bulge-to-disk mass ratios through constrains on the distributions of $\sigma_O/\sigma_S$.
Unfortunately the processes that bring stellar mass from outside a galaxy also randomize the orbits of disk stars,
and stellar mass generated {\it in situ\/} can also be transferred to bulges through secular
processes \citep{kormendy2004}. These stars remain in the bookkeeping as portions of $\sigma_S$ and $\sum_{t=1}^T S_t$,
not in $\sigma_O$ and $\sum_{t=1}^T O_t$.

The accretion of old stellar mass without substantial amounts of attendant cold gas may begin
at some late time $Q$, such that
\begin{eqnarray}
M_{n,T+1}&=\label{eq:soq}&\sum_{t=1}^TS_{n,t} + \sum_{t=Q}^TO_{n,t}
\end{eqnarray}
obtaining
\begin{eqnarray}
\mathcal{E}[M_T]&\label{eq:mothq}=&
\biggl({2\over{\pi}}\biggr)^{1/2}\biggl[
\sigma_S\biggl({T^2\over{4}}\biggr) +
  \sigma_O{(T-Q)^2\over{4}}\biggr]\\
\mathcal{E}[S_T/M_T]\label{eq:smothq}&=&2\over{T(1+B_T)}\\
B_T&=&{(T-Q)^2\over T^2}{\sigma_O\over\sigma_S}
\end{eqnarray}

Such generalizations may prove interesting from an astrophysical standpoint but are not numerically
distinct from the case where $Q=0$, as $(T-Q)^2/T^2$ is simply subsumed into $\sigma_O$. If high quality data can
uniquely specify the spectrum of $\sigma_O$, through, perhaps, measurements of the evolution of the
stellar mass function over a significantly broader range of masses than has been done to date, perhaps models
with $Q>0$ could be found useful. But showing that such a fit to the data is unique is another story (see below).

Lastly, the timescales for stochastic changes in mass accretion and star-formation may evolve separately, complicating
the mathematical story. We discuss variable timescales of stochasticity below, but if there is differential evolution
between the timescales of the
{\it in situ\/} and {\it ex situ\/} processes the results will remain mathematically
indistinguishable from the cases discussed next.

\subsubsection{Star-Formation that Declines with Time}

Generalizing the derivations to model a decline in SSFR for high-mass galaxies may also produce solutions that
fit the SFMS data.
Equation \ref{eq:so} explicitly accounts for extra stellar mass accumulation, but does not explicitly include any
process that may diminish $S$ over time. Gas mass fractions are significantly lower today
than in the past \citep{tacconi2013}. While the steady accumulation of stellar mass formed {\it ex situ\/} can lead to
lower gas mass fractions at the present epoch, the decrease may also simply mean that fuel supplies are lower.

One may then wish to revisit the question of star-formation efficiency and the
availability of gas over the long-term:
\begin{eqnarray}
M_{n,T+1}&=\label{eq:efo}&\sum_{t=1}^T\epsilon_{n,t}F_{n,t} + \sum_{t=1}^TO_{n,t}
\end{eqnarray}
where the stochastic changes may occur either to star-formation efficiencies or to the availability of fuel for
star-formation. Let us ignore {\it ex situ\/} mass growth for the moment and write
\begin{eqnarray}
M_{n,T+1}&=\label{eq:ef2}&\sum_{t=1}^T\epsilon_{n,t}F_{n,t}
\end{eqnarray}

Any expected trends in star-formation efficiency or availability of gas modify Equation \ref{eq:stationary} such that
$X_t$ is no longer a random variable centered on zero, but biased to the negative. In such cases, $X_t$ is no longer a
martingale but a supermartingale. A supermartingale is the opposite of a submartingale:
\begin{eqnarray}
\mathcal{E}[S_{t+1}]\le S_t\label{eq:super}
\end{eqnarray}
Here we apply Doob's principle that every submartingale can be decomposed into the superposition of a martingale and an
increasing sequence of nonnegative random variables \citep{hall1980}. A submartingale increases on average over time,
while a supermartingale decreases on average over time. Applying Doob's principle, we then
define $Y_t$ as an increasingly nonpositive random variable (because $S$ is a {\it super\/}martingale, not a {\it
sub\/}martingale), and write
\begin{eqnarray}
S_T&=&\sum_{t=1}^{T}X_t+\sum_{t=Q}^{T}Y_t
\end{eqnarray}
where $Q$ is a timestep at which the effectiveness of stellar mass growth begins to diminish on average.
Note that a physical interpretation of $Y_t$ may not be unique, however, as it may represent a range of processes, much
as $S_t$ can
subsume random variations in star-formation efficiency or gas inflow (\S \ref{sec:eff}).

Using this decomposition we rewrite Equation \ref{eq:ef2}, the stellar mass accumulated up to time $T$ from
star-formation as
\begin{eqnarray}
M_T&=\label{eq:superq}&\sum_{t=1}^{T}\sum_{i=1}^t X_i + \sum_{t=Q}^{T}\sum_{i=Q}^{t} Y_i
\end{eqnarray}
Because Equation \ref{eq:superq} is mathematically identical to Equation \ref{eq:soq}, we are left with
the rather disappointing conclusion that the results from encoding long-term suppression of star-formation
will look quite similar to the expectations one obtains when galaxy growth includes external sources of
stellar mass --- at least when looked at from the SFMS alone. How these impact the mass function will depend on how one
reallocates the accreted stellar mass at early times so detailed modeling of the mass function ought to break the
degeneracy.

\subsubsection{Revisiting Timescales of Stochasticity}
\label{sec:time2}

In Section \ref{sec:time} we explored simple variations in the timescales of stochasticity, whereby individual
galaxies may experience stochastic changes in SFR on their own, constant timescales. When every galaxy has its own
constant timescale, the SFMS doesn't care.

But if, as in Equation \ref{eq:dmdt2}, galaxies experience random changes to their SFR on variable timescales, the
derivation proceeds rather interestingly. Equations \ref{eq:dmdt3} and \ref{eq:dmdt4} become
\begin{eqnarray}
\langle{\dot{M}}\rangle_T
&=\label{eq:dmdt3v}&\sum_{t=1}^{T}
\bigl(\langle{\dot{M}}\rangle_t{\Delta{\mathcal{T}}_t\over\Delta{\mathcal{T}}_T}
- \langle{\dot{M}}\rangle_{t-1}{\Delta{\mathcal{T}}_{t-1}\over\Delta{\mathcal{T}}_T}\bigr)
\end{eqnarray}
and
\begin{eqnarray}
M_{T+1}&=\label{eq:dmdt4v}&\Delta{\mathcal{T}}_T
\sum_{t=1}^{T} \sum_{i=1}^{t}
\langle{\dot{M}}\rangle_i{\Delta{\mathcal{T}}_i \over \Delta{\mathcal{T}}_T}
- \langle{\dot{M}}\rangle_{i-1}{\Delta{\mathcal{T}}_{i-1}\over \Delta{\mathcal{T}}_T}
\end{eqnarray}
resulting in the expectation ($H=1$):
\begin{eqnarray}
\mathcal{E}[{\langle\dot{M}\rangle_T}/M_T]&=\label{eq:time2v}&{2\over{T\Delta{\mathcal{T}}_T}}
\end{eqnarray}
So long as galaxies at time $\mathcal{T}$ are experiencing stochastic changes in stellar mass growth on timescales that
are, on average, equal to their lifetime average timescales of stochasticity, the resulting median SSFR remains unchanged.

But if a particular set of selection criteria isolate galaxies that, on average, experience stochasticity on
timescales that are not representative of the ensemble from which the sample was drawn,
then we must write
\begin{eqnarray}
\mathcal{E}[{\langle\dot{M}\rangle_T}/M_T]&=\label{eq:time2vo}&{2\over{\mathcal{T}\mathcal{R}_T}}\\
\mathcal{R}_T&=&{\Delta{\mathcal{T}}_T\over \langle\Delta\mathcal{T}\rangle}
\end{eqnarray}

Equation \ref{eq:time2vo} has interesting consequences. One example is the Milky Way, for which \cite{snaith2014} derived a
relatively constant $\text{SFR}\approx 2-3 M_\odot/$yr for the past 7 Gyr. If that 7 Gyr timescale since the last
stochastic change in SFR is abnormally long compared to other disk galaxies of the same stellar mass, then
Equation \ref{eq:time2vo} may be a rather peculiar explanation for why the Milky Way's specific star-formation rate is
abnormally low {\it for its stellar mass\/}.

If one targets galaxies in specific regions of the universe where timescales for stochastic change in $S$
are short or long, the constituent galaxies should display biased distributions of apparent stellar mass growth
with respect to the cosmic average. Where timescales are, on average, short, the distribution of SFR at fixed mass
should skew high. In such regions, however, {\it ex situ\/} stellar mass growth will also be higher. How these
two processes are mismatched will determine whether, and for how long, galaxies sit high or low with respect to
$\langle\text{SSFR}\rangle$.

Which process wins out in shaping the observed distributions of SSFR? How does this competition shape the
star-formation-density relation \citep{cooper2007,patel2009,cooper2010,tran2010,patel2011,quadri2012}? And since
high-density regions of the universe evolve more quickly than low-density regions of the universe, do these high-density
regions simply contain distributions of galaxies with intrinsically longer timescales of stochasticity at fixed cosmic
time? Of course these environments serve to systematically suppress $S$ (or $F$) in the long-term, contributing to
systematically lower SSFR at fixed mass at high density \citep{patel2011,quadri2012}.

Note too that Equation \ref{eq:time2vo} is mathematically identical to Equation \ref{eq:smothq}. The result is that
long-term systematic lengthening of stochastic timescales will have the same mathematical consequence on
distributions of SSFR as the long-term accumulation of stellar mass from mergers and accretion, and
the same mathematical consequence on distributions of SSFR as running out of fuel for star-formation.
Thus the repetitive diminishing of SFR per unit stellar mass in high mass galaxies may arise because something occurs
to systematically lengthen the timescales over which stochastic changes in occur --- separately from processes that
diminish star-formation efficiency and/or fuel supply, or add stellar mass to galaxies from the outside.

\section{Ambiguities in Interpreting Distributions of Star-Formation Rates at
Fixed Stellar Mass}
\label{sec:ambig}

Based on the explorations above, and the correlations shown earlier between $M/L$ ratios and galaxy colors,
it appears that galaxy scaling relations, and the SFMS in particular, are degenerate integrations
of the processes that grow and shape galaxies over time. Coupling such data with, for example, the evolution of stellar
mass functions and measurements of the SFRD over time will be critical for identifying which galaxies are accreted out
of the mass function as $O$, the external source of stellar mass growth, and whether $S$ decreases with time faster
than $M$ increases. Certainly any model of the long-term evolution of galaxies must include all such processes, but at
least we now have the first step or two in assembling a meaningful statistical framework.

Even with only the beginnings of a new framework, we can already reinterpret a broad range of observations
beyond the SFMS, casting, for example, the growth of galaxies through starbursts and the rise of quiescent galaxies
in a new light.

\subsection{Predicted Starburst Fractions and Duty Cycles}
\label{sec:bursts}

Starbursts are galaxies that have rates of ongoing star-formation significantly higher than their lifetime-averaged
SFRs ($M/\mathcal{T}$). These are classified usually by $S>BM/\mathcal{T}$, where $B$ is a threshold.
Our derivations specify a distribution of SSFR at any given time for disk galaxies, and thus relates specifically
to the frequency one should expect ``starburst'' phenomena. This framework should not be interpreted in a way that implies
starburst galaxies are not real, but simply that there is a natural way to calculate the probability that galaxies,
at a given mass, have had a large fraction of their mass formed very recently (especially since growing more
rapidly would have put them into the next larger mass bin).

Equation \ref{eq:clt4h}, and Figure \ref{fig:pull4}, say that $\langle\text{SFR}\rangle$ is equal to twice the lifetime
average SFR. If one has defined a starburst as a galaxy with $\text{SFR}_\text{burst}\ge{3}\times{M/T}$, one can use
our statistical model for the distribution of SFRs at fixed mass and calculate the expected fractions of disks
that exceed such thresholds. Our fBm realizations have ``starburst'' fractions of 20\% at $t=T/2$ (half a
Hubble time), and 24\% at $t=3T/4$ (these variations belie the effects of the finite size of our fBm samples).
In other words ${20-25}\%$ of disk (bulgeless)
galaxies should meet this starburst criterion at every epoch. \cite{dressler2009}, in fact, found $\sim{25}\%$ of field
galaxies at $0.4<z<1$ in such a starburst mode. At masses where bulge mass fractions are not substantial, this starburst
fraction
should should remain relatively constant --- at least when defined relative to lifetime average SFRs. In a study of
dwarf galaxies in the local volume, \cite{lee2009} found starburst fractions of $\sim 20\%$, comparable to the
prediction here, though small number statistics and uncertain starburst criteria make a detailed comparison
more difficult.

But as bulges assemble over time, at fixed mass, this operational definition of starburst
masks the presence or meaning of burstiness, which ought to be defined in relation to disk mass.
Recall that the fraction of disks with $\text{SFR}\ge{3}\times{M_{disk}/T}$ is $\sim{20}\%$. Galaxies with bulge
mass fractions of $50\%$ would only meet the same observational threshold if $\text{SFR}\ge{6}\times{M_\text{disk}/T}$
and only $\sim 4\%$ actually meet this heightened level of activity.
Those galaxies with 75\% bulge mass fractions require even greater activity to be classified as starbursts,
at a level of $\text{SFR}\ge{12}\times{M_{disk}/T}$, and $\simlt 1\%$ of disks will meet this criterion.

The first consequence of these calculations is that as populations of galaxies acquire more and more stellar mass
from outside sources, fewer and fewer will appear as starbursts. In fact,
\cite{dressler2009} found that the starburst fraction decreased from 25\% at $0.4<z<1$
to $10-15\%$ by $z=0.3$, and $<5\%$ today. Using the typical bulge mass fractions for galaxies with
$M\sim\text{few}\times{10}^{10}M_\odot$ in the SDSS, ${1/3}-{1/2}$ \citep{abramson2014}, and
correcting the starburst criteria, one derive starburst fractions of $5-10\%$. These results
suggest that galaxies at these modest masses have acquired substantial amounts of stellar mass from outside sources,
presumably assembling it into bulges and spheroids, and the observed rapid evolution in early-type fraction at low
redshift seen by \cite{kovac2010} in the field supports such an assertion.

Disks that meet starburst definitions will only do so for a specified periods of time, and these timescales will
depend sensitively on the starburst intensity. After all, forming stars too quickly will push a galaxy into the next
mass bin, where that galaxy's SFR will be compared to an entirely different distribution of SFRs.
Our mathematical framework provides estimates for these timescales.

\begin{figure}[h]
\centerline{\includegraphics[width=2.8in]{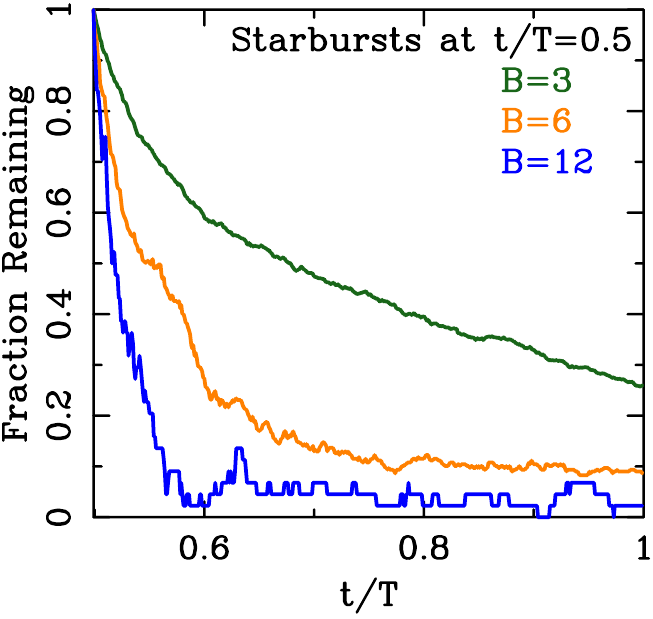}}
\caption{The duration of starbursts (green), as defined by an ongoing SFR greater than three times the lifetime average.
For pure disk galaxies, 50\% of galaxies detected as a starburst by this criterion would remain that high for about
20\% of a Hubble time, $\sim T/5$. But for galaxies that have bulge mass fractions of 50
star-formation rate $SFR>3\times M/T$ is equivalent to $SFR>6\times M_\text{disk}/T$, shown in orange. Half of these
galaxies would no longer be considered starbursts by $\sim T/20$, or about 0.5 Gyr. For galaxies with bulge mass fractions
of 75\%, this selection is equivalent to $SFR>12\times M_\text{disk}/T$, shown in blue. Such galaxies  appear as starbursts
for $\sim 100$ Myr because such extreme intensity, relative to the accumulated mass, is rare and short-lived. 
\label{fig:burst}}
\end{figure}

In Figure \ref{fig:burst} we run fBm models ($H=1$), take those models that meet starburst criteria at half a Hubble
time, and compute what fraction of those starbursts would remain classified as such at later intervals.
We calculate that of those stellar disks that have $\text{SFR}>=3\times{M_{disk}/T}$ (green line), 50\% would meet
still this threshold a fifth of a Hubble time later. As an example, more than half of the starbursting disk galaxies
at $z\sim 0.8$ would no longer be classified as starbursts by $z=0.5$ (while many other disk galaxies would be, thus
preserving the 20-25\% starburst fraction). And while 26\% of the starbursts from $t=T/2$ would still be classified
as starbursts at $t=T$, $1/3$ of these would not necessarily have been classified as starbursts during the intervening
periods of time.

But as galaxy bulges are assembled at late times the timescales over which galaxies appear as starbursts is shorter
Galaxies with bulge mass fractions of 50\% at $z=0.8$, for example, require $\text{SFR}>=6\times{M_{disk}/T}$ to be
classified as starbursts (orange line), and more than half of these starbursts  would no longer be identified as such by
$z=0.7$, $\sim 1/25$th of a Hubble time later. For galaxies with even higher bulge fractions the duration is even shorter.
Note that in each of these cases the timescales are set because the galaxy masses increase rapidly, and when
star-formation stochastically dips, the galaxies cease to be classified as starbursts. Typically this happens after
the ``starbursts'' increase the (disk) mass by $\sim 50\%$.

We remind the reader that these statistics for starbursts, and their lifetimes, are derived from the distributions of
SFRs that must be present when galaxies grow according to a stochastic process of the kind described in \S
\ref{sec:process}. Within a specific period of time, a galaxy of a given mass could have only experienced a specific set
of growth histories. The set of histories is governed by the assumption that, on average, SFRs do not change from one
timestep to the next, and the central limit theorem sets the probability that can be attached to each possible history.
The convergence, in distribution, of the long-term correlations between stochastic changes in SFR modifies these
probabilities and broadens the set of possible histories. When star-formation is not a stationary stochastic process,
the results will change, such as our simple calculations that take into account a broader range of bulge mass fractions.
Processes that grow bulges while adding to {\it in situ\/} star-formation may modify these deduced probabilities,
but only if the sum of processes cannot be reduced to, e.g., Equation \ref{eq:ff}.

Because the SFMS encodes so little information about star-formation timescales, no individual starburst can be assigned
a timescale over which it has endured such intense stellar mass growth without additional data. On average, however, the
mathematics of stochastic processes relates the duty cycles directly to the Hubble time at every epoch. For galaxies at
$z\sim{4}$, \cite{wyithe2014} estimated starburst duty cycles of $\sim{T/10}$, consistent with the predictions shown in
Figure \ref{fig:burst}.

\subsection{The Cycling of Quiescent Galaxies}
\label{sec:recycle}

Stochasticity, through the dispersion in SFHs and distribution of SSFRs at fixed mass, also provides
estimates for the fraction of disks
that are quiescent at any given epoch. Quiescence is observationally defined by the amount of ongoing star-formation
relative to the mass in old stars. To first order this is just a threshold in specific star-formation rate
\citep{williams2009,kelson2014a}. Quiescence includes galaxies with both zero ongoing star-formation, and
sufficiently low levels of relative star-formation to be observationally inconsequential. The math predicts the
fraction of systems that have $S_t=0$, and for roughly how long. In tandem with the scatter in SSFR for star-forming
systems, we can calculate the fractions of disks with low enough SSFRs to be classified as quiescent. Any disks with
appreciable bulge mass will have SSFRs that are further depressed.

Theoretically, mathematically, there are always disk galaxies with $\text{SSFR}=0$. This fraction of disks that are
effectively dead at any given time is 16\%. And 50\% of these remain dead for timescales of $\sim{T/2}$, with over 25\%
remaining so for more than half a Hubble time. In \S \ref{sec:process} we had explicitly derived the expectation values
for $H=0.5$, and in such models $\sim{1\%}$ of the population has $S_t=0$. That subset recycles very quickly, with all
of them rejoining the star forming population at every timestep. The growth histories of the dwarf galaxies, for
example, clearly all experience periods where $S_t=0$ though the sample is not large enough to statistically verify
the probabilities being generated by our framework. We save that for the next section when we explicitly compare
predicted and observed quiescent galaxy fractions.

But galaxies with zero ongoing star-formation are not the only galaxies that can be classified as quiescent.
Galaxies that appear quiescent observationally simply have low SFRs relative to their mass in old(er) stars, with a typical
criteria $\text{SSFR}\simlt 0.25/\mathcal{T}$, which we verify by propagating our SFHs through stellar population models
as in Figure \ref{fig:uvj}. At any given epoch those stellar disks that appear as quiescent do not remain so for the
remainder of history. Thus galaxies selected as quiescent at high redshift may be poor representatives of the ancestors
of galaxies deemed quiescent at lower redshifts.

\begin{figure}[h]
\centerline{\includegraphics[width=2.8in]{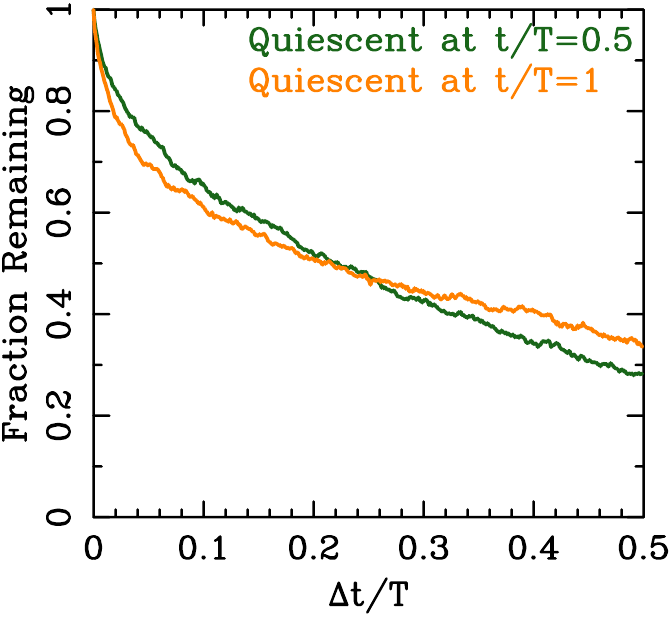}}
\caption{The duration of quiescence for disk galaxies, as defined by an $\text{SSFR}<0.25/T$.
Stellar disks that appear quiescent at half a Hubble time (green) do not remain so forever. Within 20\%
of a Hubble time only 50\% of them would still appear as quiescent. Of those that appear quiescent today (orange),
only 50\% would have been classified as quiescent 20\% of a Hubble time ago. Half a Hubble time ago, 2/3 of
those appearing quiescent stellar today would have been classified as star-forming. Such cycling of galaxies between
classifications of quiescence and star-forming have negative consequences for studies that assume quiescent galaxies at
early times comprise fair, representative samples progenitors of quiescent galaxies at late times.
\label{fig:powerspec2}}
\end{figure}

Figure \ref{fig:powerspec2} quantifies this problem. Here we show
using the green curve the fraction of (disk) galaxies selected as quiescent at half a Hubble time that would retain such
classification at later times. Within 25\% of a Hubble time more than half of the quiescent population has already
cycled back into the star-forming population. And likewise for those (disk) galaxies classified as quiescent today (but
looking back in time). The correspondence between quiescent galaxies at some distant time in the past and those today
appears to be quite poor. And these disks won't care if they have bulges --- only their aggregate SSFRs will be lower,
increasing their chances of being classified as quiescent (see below) but growing in mass nonetheless. The mathematics
for the stochastic evolution of the disk components of galaxies likely continues as derived, independent of bulge mass.

\begin{figure}[ht]
\centerline{\includegraphics[width=2.75in]{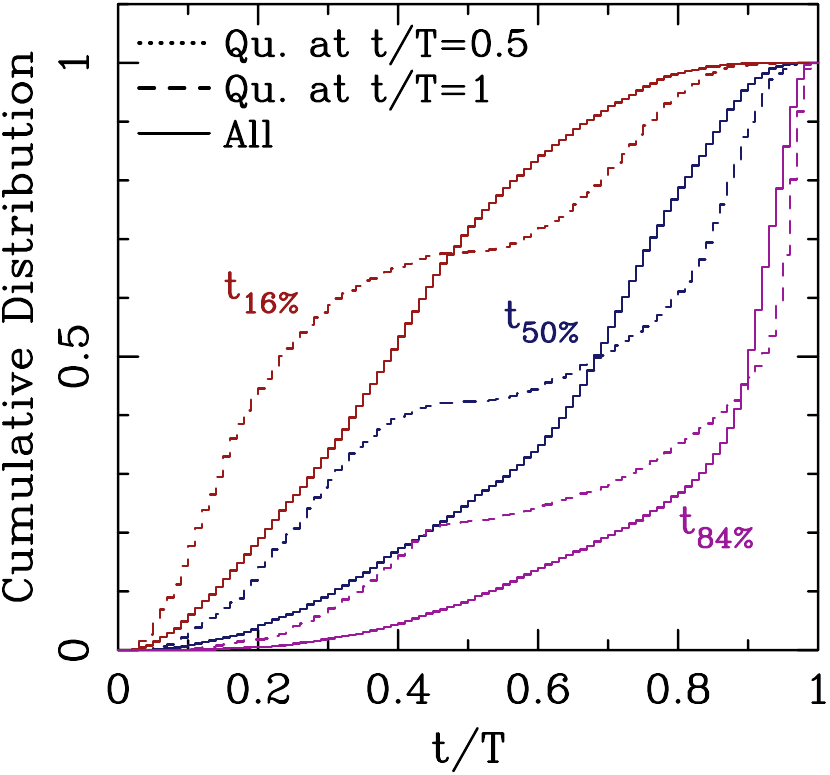}}
\caption{Cumulative distributions of the times to reach 16\%, 50\% and 84\% of the mass at time $t=T$, in red, blue, and
violet, respectively. The solid lines trace the cumulative mass evolution for disks selected as quiescent at $t=T$,
whereas the dashed lines show the continued mass evolution for disks selected as quiescent at half a Hubble time
($t=T/2$). For galaxies with significant bulge mass fractions, a larger fraction of galaxies will be counted as
quiescent but their disks will maintain distributions of mass growth similar to that shown here, with consequences
for their evolving stellar populations, galaxy sizes, and internal structures.
\label{fig:timescales}}
\end{figure}

Assuming such recycling is occurring, there are broad consequences for comparisons of the properties of ensembles
of (quiescent) galaxies  over long redshift baselines. Any analysis that requires progenitorship \cite[e.g.][and
others]{pvd2010} must fully model such recycling of galaxies. The tracing, for example, of the sizes of quiescent
galaxies over time must be seen in a context where the resumption of stellar mass growth is not only routine, but
is essentially part of the normal process by which all galaxies grow (not just the quiescent ones).
We emphasize this point in Figure \ref{fig:timescales}, in which we plot
the cumulative distributions of timescales when (disk) galaxies reach 16\%, 50\%, and 84\% of their present-day
masses (or at least the masses at the time of observation). The full distribution is shown by the solid lines.
Here we see that 50\% of these (disk) galaxies reach 50\% of their present-day mass by relatively late times, about
1/3 of a Hubble time ago. And about a quarter of the galaxies reach their 50\% mass point almost 2/3 of a Hubble time
ago. For comparison, but not shown, t050\% of (disk) galaxies selected as quiescent today reached 84\% of their
present-day mass 40\% of a Hubble time ago. Half of them reached half their present-day mass by $z\sim 0.9$.

But Figure \ref{fig:timescales} specifically shows the continued mass growth for those disks that would have
been classified as quiescent at half a Hubble time. More than half of them have not yet reached 50\% of their
present-day masses, as that won't have happened until $z\sim 0.4$.

Note that these calculations are for galaxies without bulges --- without spheroidal mass components that appear
older than the disks because the mass formed {\it ex situ\/} is
comprised of stars taken out of the continued history of stochastic mass growth.
Galaxies with (old) bulges may or may not see different SFHs for their stellar disks, depending on their situations.

Of course quiescent galaxies tend to live in rich, high-density environments at low- and high-redshifts
\citep{dressler1980,patel2009,quadri2012}. How will these galaxies continue to acquire the new material out of which
must grow new stellar mass? Even at the present epoch, only $\sim 10-20\%$ of the stellar mass density of the universe
is locked into regions that can no longer support ongoing star-formation \citep{williams2012}. In the past, galaxies
that reached states of quiescence would still have been in relatively low-mass groups, and these groups continue to
accrete and grow with time, bringing fresh fuel to rejuvenate stellar disks \cite[e.g.][]{wetzel2013}.

Thus all studies of the evolution of quiescent galaxies over time must account for (1) mechanisms that stochastically
reignite star-formation in the old disks of quiescent galaxies, presumably in disks, or (2) recent stochastic emergence
of quiescent galaxies out of those populations of galaxies that were recently forming stars. Assuming some mean mass growth
\citep[e.g.][]{leitner2012} does not mitigate against these issues.

\begin{figure*}[t]
\centerline{\includegraphics[width=7.0in]{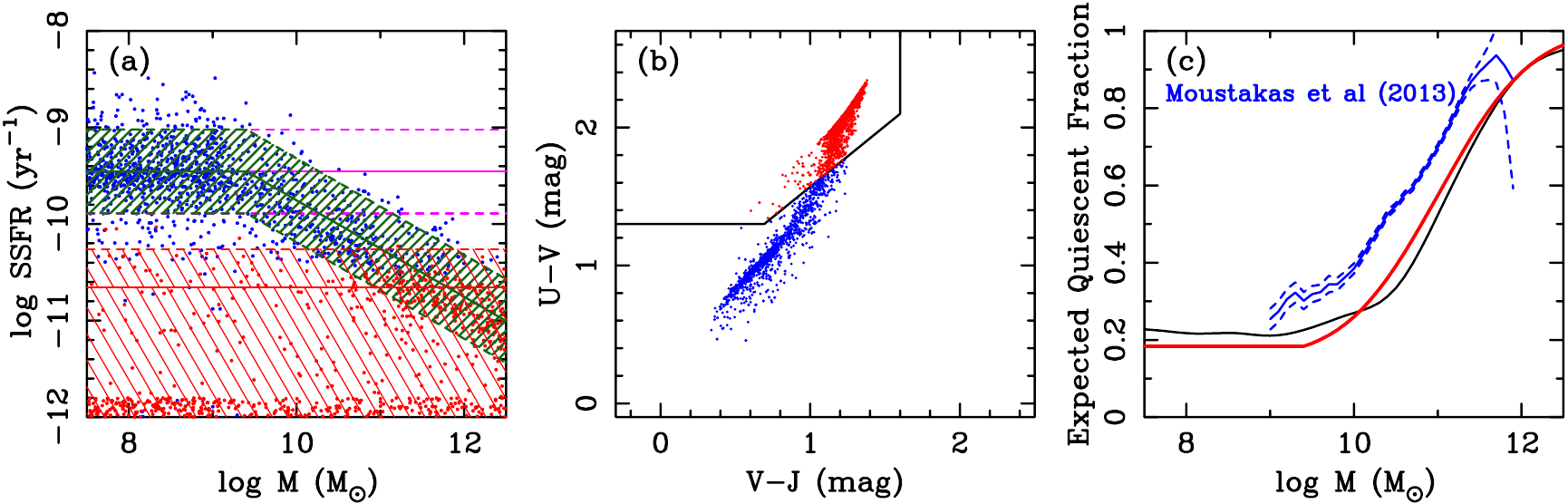}}
\caption{(a) Schematic of the SFMS and the selection of quiescent galaxies at $z\sim 0.1$.
The blue solid and dashed lines show predicted median SSFR and its scatter of $\sim 0.4$ dex ($H=1$).
The green locus applies the broken power-law form
from \cite{salim2007}, which shows a roughly flat $\text{Median}[\text{SSFR}]$ below $log M\le 9.4$ and a decline of
$\sim -0.5$ dex/dex at higher masses.
The selection of quiescence is roughly equivalent to the solid red line, though the mean attenuation
of $A_V\sim{1}$ mag for star-formation means that galaxies up to the red dashed line, on average, will
also be classified as quiescent. Details of the observational selection of quiescence will also alter this
selection, as the points were color coded red and blue according to their $U-V$ and $V-J$ colors.
(b) Model $U-V$ vs $V-J$ bicolor diagrams for the model SFHs that are shown in (a). The points are colored red and
blue for those that fall within the \cite{williams2009} boundaries separating quiescent from star-forming galaxies.
(c) Using the volume under the lognormal defined by the green hatched regions, plus
$\sim 16\%$ of the population with $\text{SSFR}\equiv 0$ (when $H=1$), we calculate
the expected quiescent fraction as a function of stellar mass shown by the red line. These simple retrodictions are
overplotted against quiescent galaxy fractions from SDSS \citep{moustakas2013} in blue and match surprising well
considering these calculations did not model the stellar populations, dust attenuations, and survey sensitivities to
star-formation activity in any way. Modeling the stellar populations, deriving restframe colors, and selecting galaxies
as quiescent based on $UVJ$ color criteria produces the black solid line.
\label{fig:sloan}}
\end{figure*}

\begin{figure*}[t]
\centerline{\includegraphics[width=5.0in]{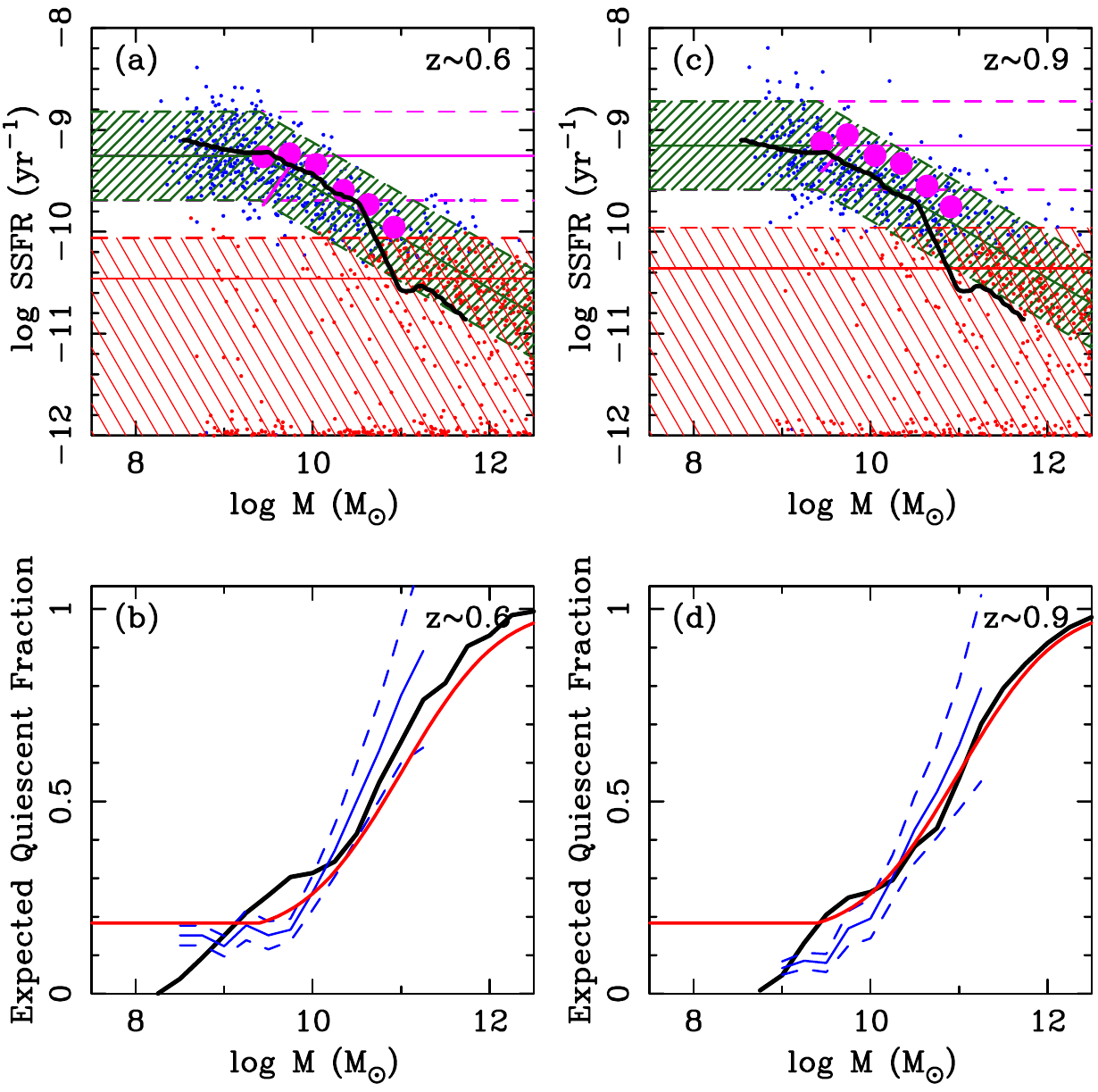}}
\caption{(a,c) Schematic of the SFMS and the selection of quiescent galaxies at $z\sim 0.6$ and $z\sim 0.9$.
The blue solid and dashed lines show predicted median SSFR and its scatter of $\sim 0.4$ dex ($H=1$).
The green locus applies the broken power-law form
from \cite{salim2007}, which shows a roughly flat $\text{Median}[\text{SSFR}]$ below $log M\le 9.4+0.5\log\mathcal{T}$
and a decline of $-0.5$ dex/dex at higher masses. Blue and red points symbolize model SFHs that are selected
as star-forming or quiescent based on their $U-V$ and $V-J$ colors at the indicate redshift. Acquisition of stellar mass
from outside sources is included in the models to reproduce the desired trend of SSFR at high masses, where the SFHs of
the acquired stellar mass at every epoch until the time observations are drawn from the random distribution of
stochastic histories. The $K$-band limit of Z-FOURGE of $K_{AB}=25.3$ \cite{tomczak2014} is included as a selection
criteria for the models. The violet filled circles are the data from \cite{karim2011} showing the median SSFR at a given
stellar mass from radio observations. The black solid lines show the trends of $\text{Median}[\text{SSFR}]$ with stellar
mass published by \cite{kajisawa2010}.
(b,d) We calculate the expected quiescent fraction as a function of stellar mass shown by the red line in the
same fashion as in Figure \ref{fig:sloan}. Modeling the stellar populations, deriving restframe colors, and selecting
galaxies as quiescent based on $UVJ$ color criteria produces the black solid line. These retrodictions are
overplotted against quiescent galaxy fractions from Z-FOURGE \citep{tomczak2014} in blue and match surprising well.
Note that the selection limit appears to reduce the quiescent fractions at low mass owing to the strong dependence of
magnitude on both stellar mass and SSFR.
\label{fig:zfourge}}
\end{figure*}

\subsection{The Dispersion in SSFR and its Consequences for Quiescent Galaxy Fractions}
\label{sec:abundance}

The martingale central limit theorem remains valid even when superposed on long-term trends. We discussed some potential
long-term trends that would add and subtract from our predictions of stellar mass growth earlier, but choose for the
moment not to fit any data to constrain such long-term processes --- partly because doing so is beyond the scope of
this purely theoretical paper, and partly because so much of the mathematics appear to be degenerate.

In this section we recognize that there is a significant change in the slope of the SFMS at high masses
\cite[e.g.][]{salim2007,whitaker2012,sobral2014}, and we simply superimpose our statistical framework onto the observed
change in slope at high mass. This is functionally equivalent to incorporating the relevant processes to recover the
proper form of the SFMS. For simplicity we adopt the broken power-law form found by \cite{salim2007}, but assume that
below $\log M=9.4$, $\text{Median}[\text{SSFR}]$ is a constant defined by our fBm models with $H=1$, and above that mass
there is a slope of $-0.5$ dex/dex. We also adopt the scatter derived with $H=1$, knowing that the central limit
theorem remains operable when superposed on long-term expectations. We expect that an additional process, or additional
processes, have produced the anticorrelation at high masses and the dispersion in the ensemble is expected to remain
$\text{Sig}[\text{SSFR}]=\mathcal{E}[\text{SSFR}]$.

The baseline SSFR and scatter are shown in Figure \ref{fig:sloan}(a) using the violet horizontal lines. The
broken power-law form is shown in green. Using only this diagram, we compute quiescent fractions as a function of mass.
Later we will discuss how this change in slope may arise in the context of our derived equations.

To first order, all galaxies below the solid red line (SSFR $<0.25/T$) are quiescent. And if additional star-formation
is attenuated by $A_V=1$ mag, those galaxies may rejoin the quiescent clump within $10^8$ yr \citep[at least in a $UVJ$
diagram;][]{patel2011}. Thus there is an effective quiescence cut shown by the dashed red line whereby low levels of
recent, attenuated star-formation are inconsequential. When superposed on galaxies with bulges, such star-formation is,
fractionally, even less consequential \citep{abramson2014}. We have propagated the fBm SFHs, including a secondary
process of {\it ex situ\/} mass growth to mimic the departure from $\text{Median}[\text{SSFR}]$ at high masses, through
stellar population synthesis models. The resulting $U-V$ and $V-J$ bicolor distribution is shown in Figure
\ref{fig:sloan}(b), assuming solar metallicity and no dust attenuation, and, despite these shortcomings the distribution
looks similar to those found in existing surveys \cite{williams2009,whitaker2012,tomczak2014,kelson2014a}. Using the
\cite{williams2009} boundaries to select quiescent from star-forming, we compute the quiescent galaxy fraction as a
function of stellar mass, shown in Figure \ref{fig:sloan}(c) by the black solid line. Just using the predicted lognormal
scatter, plus the 16\% of galaxies with $S_T=0$ at $z_\text{obs}$ (see \S \ref{sec:recycle}), one can estimate a rough
trend of quiescent fraction with galaxy stellar mass, shown by the red line in Figure \ref{fig:sloan}(b). The observed
quiescent galaxy fractions from SDSS \citep{moustakas2013} are shown in blue and match surprising well, though the
\cite{moustakas2013} definition of quiescent was defined using color-magnitude relations. Because we have no dust
attenuation in our simple tests, we have opted to rely on the $UVJ$ criteria. Furthermore, at the present epoch, $\sim
10\%$ of the stellar mass density of the universe is locked into regions that can no longer support ongoing
star-formation \citep{williams2012} and these simple models have not enhanced the quiescent fractions to account for the
increased volume comprised of such evolutionary dead-ends.

Despite these shortcomings, the model works surprisingly well. When comparing to higher redshift data
some of the more problematic issues are mitigated, such as the diminished contributions of the richest groups
and clusters to the global average, and the more common selection of quiescent galaxies using $UVJ$ diagrams.
So in Figure \ref{fig:zfourge} we perform the same exercise at redshifts $z\sim 0.6$ and $z\sim 0.9$, making direct
comparisons to the quiescent galaxy fractions in Z-FOURGE \cite{tomczak2014}. Figures \ref{fig:zfourge}(a,c) show the
SFMS, with the assumed anticorrelation at high masses. Overlayed are data from \cite{karim2011} and \cite{kajisawa2010},
to illustrate that the model SFMSs are reasonable starting points. Simple integrations over the underlying dispersion in
SSFR are shown by the red lines in Figures \ref{fig:zfourge}(b,d), and are analogous to that shown in Figure
\ref{fig:sloan}(c).  But to better match the published data, we have constructed magnitudes in the observed frame,
``selecting'' the samples to be limited at $K_{AB}=25.3$ mag, the limit of the Z-FOURGE survey. The resulting quiescent
galaxy fractions are shown by the black lines, which closely mimic the observed quiescent fractions shown in blue.

While careful modeling of the observational errors, dust attenuations, stellar abundances, and selection effects may
improve the overall quality of the mimicry, the basic match between the model expectations and the data is striking.
Such estimations call into question the validity of separating galaxies into categories based on quiescence, and we
suggest that a broader view be taken when modeling the evolving distributions of (all) galaxies over time.

These results should not be taken to imply that there are no quiescent galaxies in the universe. Just as in the
discussion of starburst galaxies, these results suggest that these classifications oversimplify the evolution of
the ensembles. Quiescent galaxies are a heterogeneous population, arising from a range of histories. Processes
that stifle star-formation in one set perhaps should not be assumed to have been the dominant mechanisms that
lead other galaxies to appear as quiescent. Earlier we discussed several avenues for expanding
the formalism of stochastic processes, including ones that can increase a the probabilities for quiescence
over time: {\it ex situ\/} mass growth, variable timescales for stochastic changes in {\it in situ\/} mass growth,
and diminished fuel supplies or star-formation efficiencies. Any of these may be operable, but the underlying
astrophysical causes are sufficiently different that one would not wish to put those galaxies that appear
quiescent because of each process into a single category. Additional data must be used to better constrain the
mechanisms that drive relative quiescence, such as the utilization of bulge mass fractions by \cite{abramson2014}.
Monolithic probabilistic formalisms are very likely not helpful, but a wealth of data regarding galaxy environments,
star-formation rate densities, structural parameters, and galaxy mass functions, should all be relied upon.

We invoked no mechanisms to permanently quench galaxies or make them appear quiescent, outside of (1) whatever modest
astrophysical events stochastically change star-formation over time, and (2) whatever process imposes
the modest anticorrelation of SSFR with galaxy mass at the high-mass end of the mass function. At least part of this
change in slope is due to an increase in bulge mass, systematically depressing SSFR so that one under-appreciates
the extent to which those galaxies are continuing to grow.

The simplest model for star-forming disks invokes no stochastic changes to star-formation
over time, producing long-term expectations of $\text{SSFR}=1/T$ with zero intrinsic scatter. This is
mathematically equivalent to $H=0$ and functionally equivalent to the models of \cite{peng2010}. In such a model no
galaxies ever reach quiescence without the invocation of additional mathematical tools or probabilistic approaches to
galaxy quenching. The reader may conclude from all these derivations that we have merely rederived the
\cite{peng2010} approach, but with enough moments in the calculations to avoid making quenching fraction a model
parameter. From a distance this interpretation may be appealing, but it would miss (1) the salient point that galaxies
only stochastically ``quench'' temporarily, and (2) that no adoption of a universal star-formation efficiency is
required or even relevant. Furthermore, one can very nearly calculate the quiescent galaxy fraction at a given
stellar mass with almost no data whatsoever.

Without even knowing any of the critical mechanisms, a shocking amount of the statistical properties of galaxies and
their evolution with time has now been deduced and we can only rather anticlimactically conclude that the rise of
quiescent galaxies at late times was largely inevitable --- the result of modest dispersion in SSFR at fixed mass coupled
with an anticorrelation of SSFR with $M$ at high masses.

\section{Conclusions}
\label{sec:conclusions}

Statistical distributions of astronomical objects offer the most insight into astrophysical events, timescales, and
processes, and have played this role for nearly a century \cite{hubble1926}. Often there are correlations between
properties, such as with galaxy sizes, luminosities, characteristic velocities, line absorption, and others
\citep[e.g.][]{fj76,tully1977,dd87,dress7s,terlevich}. The SDSS has revolutionized such studies by ballooning
collective dataset of galaxies by factors of many thousands \citep[e.g.][]{kauffmann2003,blanton2005}.

Out of such large samples emerge patterns such as the correlation between on-going rates of star-formation with galaxy
mass \citep{brinchmann2004,salim2007}. Such results appear when a practically infinite number of objects have been
averaged to reveal mean behaviors and properties. But by averaging the properties of an infinite number of
objects we are using the Universe as an analog computer to derive the asymptotic behavior of an ensemble,
top compute the limiting behavior of such properties in distribution. So when such distributions are observed, and mean
behaviors are revealed, do they mean nothing? The work presented in this paper provides an ambiguous answer.

To zeroth order the existence of a correlation between SFR and $M$ indeed means very little, as a
number of theories and frameworks have been published that non-uniquely construct such a correlation. That the scatter
in the SFMS is so low has led many to draw significant conclusions about how {\it individual\/} galaxies form and grow,
leading to quite sophisticated models \citep{peng2010}. In such frameworks, powerful quenching mechanisms have been
postulated, and required, to explain the rise of quiescent galaxies over time
\citep[e.g.][]{peng2010,peng2012,behroozi2013}.

Our work, however, has shown that the limiting behavior of ensembles cannot be used to infer the paths of individual
galaxies. But in deriving basic properties of the ensembles of star-forming galaxies from first principles, we have
uncovered what is essentially a set of nondifferentiable basis functions with which one can model galaxy SFHs. With each
of these SFHs even come probability densities, though our framework clearly requires additional pieces to fully model
galaxy assembly, let alone the assembly of their dark matter halos.

Simple, deterministic frameworks for galaxy evolution are understandably attractive, even --- or especially --- in the
face of chaotic cosmological simulations. Such frameworks make for convenient equations, with terms that appear to
confirm intuition and provide insight. Without such relatively simple mathematical frameworks, it has been difficult
to make progress deciphering galaxy distributions over cosmic time. Fortunately, 20th and 21st century
mathematical tools are available to better comprehend the evolving
distributions of galaxies, though the differences between the philosophy that underpins this new work and competing
analytical and semi-analytical frameworks could not be more striking.

With no assumptions about star-forming efficiency or gas fractions, central limit theory for stochastic processes yields
a star-forming main sequence that agrees with the SSFRs of (star-forming, low-ish-mass) galaxies over $0<z<10$. Our
derivations show that their median SSFR declines as $\mathcal{T}^{-1}$, is independent of disk stellar mass, is
independent of the timescale for stochastic changes in stellar mass growth, and does not depend on a universal
efficiency of star-formation. Ultimately  --- because galaxies ``remember'' what they've done before --- one parameter
was introduced that controls the aggregate distribution of covariance, but its value is itself highly constrained by the
data to be $H=1$, both through the differences between published mean and median SSFRs, and the definition that median
SSFR scales as $(1+H)/\mathcal{T}$. There is real intrinsic dispersion in
$\log\text{SSFR}$ of $\sim 0.3-0.35$ dex, depending on the method of measurement and on the nature of one's sample
selection.

And while such a model is extreme in its unattractiveness, it has a precision and accuracy that is unparalleled in its
lack of free parameters. The accuracy with which the retrodictions and data agree implies that the SFMS can be used to
measure cosmological parameters, with data from the literature constraining the Hubble constant to 5\% (random, and 7\%
systematic).

Constructing a method to turn star-forming galaxies into standard clocks was not the intent of the work, but
the agreement between the predictions and the data has broad implications for the underlying SFHs of galaxies in
general. We summarize some of the key points here:

{\parindent=0pt\parskip=2pt
$\bullet$ A stochastic process is not anarchic, but is expected to remain unchanged from one epoch to the next. This
basic assumption appears to be foundational for constructing SFHs.

$\bullet$ The star-formation histories of dwarf galaxies \citep[e.g.][]{skillman2014,weisz2014}, and the Milky Way
\citep{snaith2014}, are consistent with SFHs generated by a stochastic process.

$\bullet$ There is a distribution in SSFRs at any epoch, and at the present epoch this distribution produces a scatter
in $M/L$  ratios consistent with the scatter in the Tully-Fisher relation \citep{pizagno2007}.

$\bullet$ The evolution of the shape of the stellar mass function is naturally explained by the 
time-dependence of the expectation values for SFR and stellar mass, such that mass evolution does not keep up with the
increasing linear size of logarithmic mass bins. This is a primary reason why the late-time double \cite{schechter1976}
function naturally transitions to a single \cite{schechter1976} function at high redshift.

$\bullet$ Because starbursts are defined in relation to lifetime average SFRs, their abundances and lifetimes are
straightforwardly calculated, and agree with published data

$\bullet$ Star-forming disks stochastically stop and restart on timescales of half a hubble time, making the assignment
of progenitorship difficult when analyzing the evolution of galaxy properties

$\bullet$ The population of quiescent galaxies (at least at late times) is heterogeneous and comprised of
galaxies that have not permanently quenched. Many have SFRs that are only temporarily low, though those that were
quenched and have zero on-going star-formation may stay that way for half a Hubble time.

$\bullet$ The modest dispersion in SSFR at fixed mass, and the anticorrelation of SSFR with galaxy mass at high masses,
naturally produces the trend of quiescent galaxy fractions with stellar mass.
}

Even though these conclusions can already be drawn from these first steps towards understanding the statistical
distributions of galaxies, the modeling does not yet include {\it ex situ\/} sources of stellar mass, or any
attendant treatment of the host dark matter halos. Furthermore, no attempt has been made to connect particular stochastic
changes in $S$ to astrophysical processes. The central limit theorem has so far been applied in a vacuum, while
simulations and galaxy formation theory may, for example, allow us to restrict which fBm paths are potentially
unphysical. With a more complete model, however, we expect to be able to derive plausible, global probability densities
for the broad range of possible galaxy formation histories at each epoch. Such distributions might then allow for the
construction of more complete sets of plausible SED templates for photometric redshift surveys with sensible priors, for
example. At a minimum, however, all astrophysical processes should be seen to sit on top of the derivations
and predictions made here.

While there are other ways to model correlations of SFR with galaxy mass over cosmic time, each requires
direct fits of parametrized functions to observed star-formation rates and stellar mass functions
\citep{behroozi2013}, or by inferring star-formation histories from the evolution of the SFRD and
distributions of star-formation rates \citep{gladders2013}. None of these constraints should necessarily be seen as
orthogonal, as they are tackling different aspects of the same problem. It should be noted that nowhere did we derive a
mass function from first principles --- only the SFMS.

The downside is that all governing physics has been shifted to spectra of perturbations, where the future histories of
$\sigma_{n,t}$'s ultimately determine what range of masses an object may attain in the future. Perhaps the matter power
spectrum, merger trees, and the predicted evolution of the halo mass functions will provide a basic framework for
calculating the spectrum of $\sigma$'s but such efforts are beyond the scope of this paper.

Other frameworks have also been made to agree with data \citep{peng2010,behroozi2013}. But free of any fine-tuning, the
agreement between our mathematical framework and observations implies that the growth of (disk) galaxies is a stochastic
process, in the mathematical sense. The SFMS is therefore not a deterministic law, dictating that galaxies have higher
star-formation rates as they grow. Galaxies at fixed mass exhibit a range of star-formation rates only because of the
central limit theorem. In other words, the SFMS is {\it descriptive\/}, not {\it prescriptive\/}.

A model with no physics and no determinism is not attractive, but a surprising amount about galaxy evolution emerges
anyway. How can this be? To first order, the answer lies in the fact that most galactic observations of astrophysical
and cosmological interest are deeply connected to SSFR --- colors, $B/D$ ratios, $M/L$ ratios, absorption and emission
line equivalent widths. If all of our observations are ratios of property $W$ to $\int Wd\text{t}$ or some suitable mapping thereof,
how do we proceed to learn anything when their behavior in distributions are the result of central limit theorems?
Perhaps ancillary data, such as structural parameters and direct measurements of bulge and disk sizes or masses will
provide helpful clues \citep[e.g.][]{abramson2014}. But the interpretation of such results, again, may not be unique:
are galaxies in high-density environments losing their disks? failing to regrow disks? growing their bulges?
experiencing longer timescales between stochastic events? This last option is a particularly novel outcome of the
derivations and should be explored further, given that stochastic change likely occurs on the dynamical timescales of
one's halo. Naturally, satellites must experience stochasticity on timescales defined by the halos of their hosts,
compared to galaxies of the same mass that reside in their own halos.

Large, complete, statistical modeling may resolve these degeneracies, but only if one begins with a sensible statistical
framework. The one presented here is not yet complete, as the mathematics for stochastically driving bulge growth,
starving fuel supplies, or lengthening timescales of stochasticity have not been fully incorporated. Nor has this model
been placed in a modern cosmological context. In this framework, how do dark matter halos grow? The growth of stellar
mass is particularly chaotic and the large dispersion in fractional mass growth at any given time quickly erases any
one-to-one mapping of halo mass to stellar mass from one timestep to the next --- especially when timescales may also
vary randomly, or at least remain unencoded in the SFMS.

The SFMS is thought of as a scaling relation, but it is correlations between a time-variable property and its integral.
Other such correlations may be reducible in a similar fashion, such as (1) the long-term evolution of the zeropoint,
slope, and scatter for the mass-metallicity relation for galaxies
\citep{lequeux1979,tremonti2004,erb2006,wuyts2012,wuyts2014}; (2) the long-term evolution of
the zeropoint, slope, and scatter for the correlation between $\text{[N/O]}$ and $\text{[O/H]}$ for galaxies
\citep{kobulnicky1999,pilyugin2004a,pilyugin2004b};
(3) the dispersion in Eddington ratios for AGN, whereby one is also taking the ratio of a mass growth rate
to the integral of that rate \citep{kollmeier2006}, or (4) the slope and scatter of Kennicutt-Schmidt-style correlations
\citep{kennicutt1998b}, among many other knotty topics.

Pessimists among us may push the most deeply unsettling aspects of this work to the extreme, whereby the dominant
processes that shape and grow galaxies may never be satisfactorily resolved. Perhaps there are profound limits to what
we may truly learn from the distributions of galaxy properties over cosmic time. Certainly this work should provide a
cautionary tale, that statistical astronomy may not have led us to an oasis of knowledge but merely
to an attractive mirage. And perhaps we are only a quarter of the way through our forty years.

Some readers may be concerned that abandoning deterministic laws is tantamount to dismissing galaxy study as
futile. Our conclusions do not fit neatly into the dominant analytical paradigms, and humans have been searching
for the natural laws that govern the heavens for quite a while. However, we may have merely, finally recognized the
deeply statistical nature of the problems, and perhaps ever more powerful mathematical tools will lead to greater,
or more nuanced, understanding. After all, underpinning all the mathematical chaos in galaxy evolution are the
astrophysical processes of ram-pressure stripping, tidal interactions, accretion, AGN, winds, and supernovae feedback,
reionization, hot halos, etc, and these all remain critically important foci of study. One should now think of
these as processes that drive stochasticity, but perhaps observationally as processes that
move galaxies around the diagrams of SSFR and mass, moving them around with respect to expectation values at each epoch,
and for specific timescales. With modern mathematical tools, such as the martingale central limit theorem, we have only
taken the first steps in what are clearly going to be new, exciting opportunities for investigation.

\acknowledgments

L. Abramson and A. Dressler are grudgingly thanked for talking too loudly in the hallway about the SFMS. A. Benson is
thanked for giving early drafts a look. Detailed editing by R. Powers made the text more palatable, and A. Dressler
provided helpful suggestions for to ensure a broader reach. B. Mandelbrot is also posthumously thanked for bringing
Kolmogorov's fBm models to a wider audience. Gratitude is expressed to R.~J. Williams for seeing early on that this work
might not be totally bogus. For tolerating the kinds of diversions that lead down such unexpected rabbit holes: deep
gratitude to the Carnegie Institution. No animals were harmed in the rendering of this work.

\end{document}